\documentclass[aip, reprint]{revtex4-1}
\usepackage{color}
\usepackage{graphicx}
\usepackage{amsfonts}
\usepackage{amsmath}
\usepackage{amssymb}
\usepackage{braket}
\usepackage{mathtools}
\usepackage{siunitx}
\usepackage{tablefootnote}
\usepackage{calc}
\usepackage{verbatim}
\usepackage{textcomp}
\usepackage{scrextend} 
\usepackage[pdftex,
pdfauthor={Matteo Mariantoni},
pdftitle={Thermocompression Bonding Technology for Multilayer Superconducting
Quantum Circuits},
pdfkeywords={Quantum Computing; Scalable Qubit Architectures; Superconducting
Qubits; Thermocompression Bonding; Flip Chip; Wafer Bonding; Indium Films;
Three-Dimensional Wiring; Quantum Socket; Multilayer Microwave Integrated
Quantum Circuits; Microfabricated Air Bridges; Through-Silicon Vias}]{hyperref}

\DeclarePairedDelimiterX{\abs}[1]{\lvert}{\rvert}{\ifblank{#1}{{}\cdot{}}{#1}}
\DeclareSIUnit\gauss{G}

\begin{document}

\preprint{}

\title{Thermocompression Bonding Technology for Multilayer Superconducting
Quantum Circuits}

\author{C.R. H.~McRae}
\affiliation{Institute for Quantum Computing, University of Waterloo, 200
University Avenue West, Waterloo, Ontario N2L 3G1, Canada}
\affiliation{Department of Physics and Astronomy, University of Waterloo, 200
University Avenue West, Waterloo, Ontario N2L 3G1, Canada}

\author{J. H.~B\'{e}janin}
\affiliation{Institute for Quantum Computing, University of Waterloo, 200
University Avenue West, Waterloo, Ontario N2L 3G1, Canada}
\affiliation{Department of Physics and Astronomy, University of Waterloo, 200
University Avenue West, Waterloo, Ontario N2L 3G1, Canada}

\author{Z.~Pagel}
\thanks{Present address: Tufts University, Department of Physics and Astronomy,
574 Boston Avenue, Medford, Massachusetts 02155, USA}
\affiliation{Institute for Quantum Computing, University of Waterloo, 200
University Avenue West, Waterloo, Ontario N2L 3G1, Canada}

\author{A. O.~Abdallah}
\affiliation{Institute for Quantum Computing, University of Waterloo, 200
University Avenue West, Waterloo, Ontario N2L 3G1, Canada}
\affiliation{Department of Physics and Astronomy, University of Waterloo, 200
University Avenue West, Waterloo, Ontario N2L 3G1, Canada}

\author{T. G.~McConkey}
\affiliation{Institute for Quantum Computing, University of Waterloo, 200
University Avenue West, Waterloo, Ontario N2L 3G1, Canada}
\affiliation{Department of Electrical and Computer Engineering, University of
Waterloo, 200 University Avenue West, Waterloo, Ontario N2L 3G1, Canada}

\author{C. T.~Earnest}
\affiliation{Institute for Quantum Computing, University of Waterloo, 200
University Avenue West, Waterloo, Ontario N2L 3G1, Canada}
\affiliation{Department of Physics and Astronomy, University of Waterloo, 200
University Avenue West, Waterloo, Ontario N2L 3G1, Canada}

\author{J. R.~Rinehart}
\affiliation{Institute for Quantum Computing, University of Waterloo, 200
University Avenue West, Waterloo, Ontario N2L 3G1, Canada}
\affiliation{Department of Physics and Astronomy, University of Waterloo, 200
University Avenue West, Waterloo, Ontario N2L 3G1, Canada}

\author{M.~Mariantoni}
\email[Corresponding author: ]{matteo.mariantoni@uwaterloo.ca}
\affiliation{Institute for Quantum Computing, University of Waterloo, 200
University Avenue West, Waterloo, Ontario N2L 3G1, Canada}
\affiliation{Department of Physics and Astronomy, University of Waterloo, 200
University Avenue West, Waterloo, Ontario N2L 3G1, Canada}


\date{\today}

\begin{abstract}
Extensible quantum computing architectures require a large array of quantum
devices operating with low error rates. A quantum processor based on
superconducting quantum bits can be scaled up by stacking microchips that each
perform different computational functions. In this article, we experimentally
demonstrate a thermocompression bonding technology that utilizes indium films as
a welding agent to attach pairs of lithographically-patterned chips. We perform
chip-to-chip indium bonding in vacuum at~\SI{190}{\degreeCelsius} with indium
film thicknesses of~\SI{150}{\nano\meter}. We characterize the dc and microwave
performance of bonded devices at room and cryogenic temperatures.
At~\SI{10}{\milli\kelvin}, we find a dc bond resistance
of~\SI{515}{\nano\ohm\per\milli\meter\squared}. Additionally, we show minimal
microwave reflections and good transmission up to~\SI{6.8}{\giga\hertz} in a
tunnel-capped, bonded device as compared to a similar uncapped device. As a
proof of concept, we fabricate and measure a set of tunnel-capped
superconducting resonators, demonstrating that our bonding technology can be
used in quantum computing applications.
\end{abstract}


\maketitle

The field of quantum computing~\cite{Ladd:2010} is experiencing major growth
thanks to the development of architectures with ten or more quantum
bits~(qubits).~\cite{Monz:2011, Kelly:2015} The biggest challenge in the
realization of a universal quantum computer is the implementation of extensible
architectures where qubit operations can be performed with low error
rates.~\cite{Martinis:2015} Among many promising qubit
architectures,~\cite{Cody:2012, Monroe:2013, OGorman:2016, Lekitsch:2017} those
based on superconducting quantum circuits~\cite{Clarke:2008} are rapidly
reaching a level of maturity sufficient to demonstrate supremacy of a digital
quantum computer over the state-of-the-art classical
supercomputer.~\cite{Boixo:2017} Elements of quantum-error correcting
codes~\cite{Gottesman:2010, Fowler:2012} have already been demonstrated in a
variety of experiments using superconducting qubits~\cite{Corcoles:2015,
Riste:2015, Kelly:2015} and a quantum memory has been realized with quantum
states of microwave fields.~\cite{Ofek:2016}

\begin{figure}[t!]
	\centering
	\includegraphics[width=0.495\textwidth]{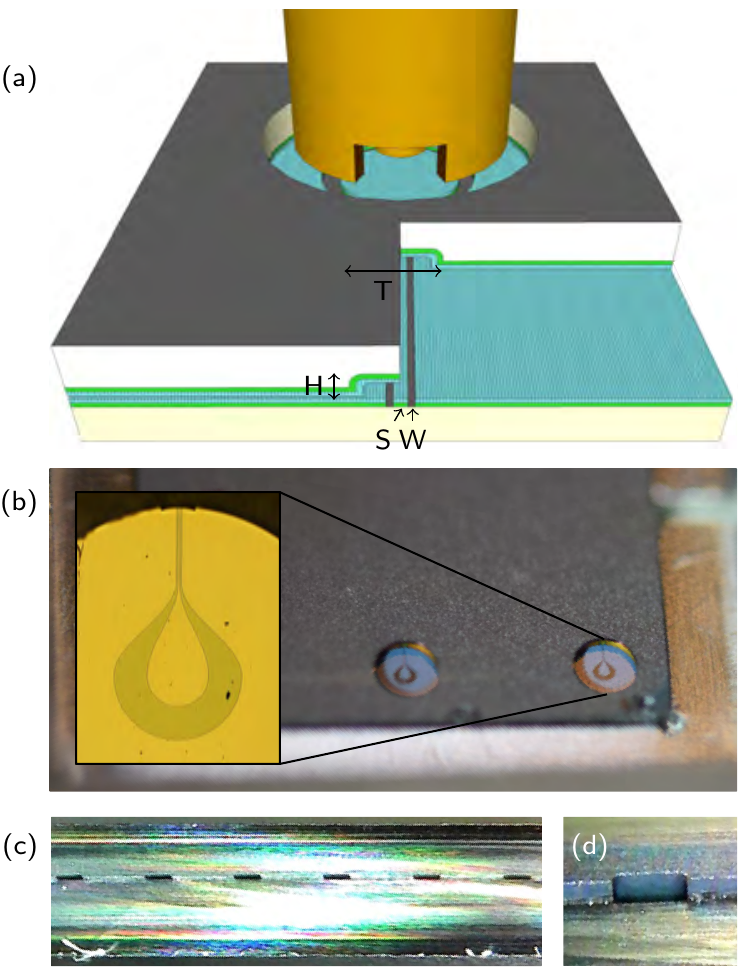}
\caption{Capped device formed using thermocompression bonding. (a) Cutaway of a
capped device, exposing aluminum [green (dark gray)] and indium [sky blue (light
gray)] films. The CPW transmission line features a center conductor of width~$S
= \SI{12}{\micro\meter}$ and gaps of width~$W = \SI{6}{\micro\meter}$. The
tunnel height is~$H = \SI{20}{\micro\meter}$, with width~$T =
\SI{175}{\micro\meter}$. Through holes in the cap allow electrical connection to
the base chip by means of three-dimensional wires. (b) Macrophotograph of a
capped device. Inset: Microimage of a through hole showing a conductor trace
aligned with a tunnel. (c) Cross-section of a capped device showing six tunnels
(dark gray rectangles). (d) Detail of a tunnel in~(c).}
	\label{Figure1:McRae:Main}
\end{figure}

In order to build an extensible quantum computer, however, many technological
advances must first be demonstrated. Among these, three-dimensional integration
and packaging of superconducting quantum circuits is emerging as a critical area
of study for the realization of larger and denser qubit architectures. This
approach allows the departure from the two-dimensional confinement of a single
microchip to a richer configuration where multiple chips are overlaid.
Three-dimensional integration, thus, provides a flexible platform for more
advanced classical manipulation of qubits and qubit protection from the
environment. In this framework, an architecture based on multilayer microwave
integrated quantum circuits has been proposed~\cite{Brecht:2016} and some of its
basic elements realized, showing that high-quality micromachined
cavities~\cite{Brecht:2015} can be used to implement three-dimensional
superconducting qubits.~\cite{Brecht:2017} Leveraging the extensive body of work
developed in the context of classical integrated circuits, flip chip technology
has been adopted to bond pairs of microchips containing superconducting
circuits.~\cite{OBrien:2017, Rosenberg:2017, Kelly:2017} Microfabricated air
bridges have been utilized to reduce on-chip electromagnetic
interference.~\cite{Chen:2014} High-frequency through-silicon
vias~\cite{Versluis:2016} and a quantum socket based on three-dimensional
wires~\cite{Bejanin:2016} have been developed to attain dense connectivity on a
two-dimensional array of qubits.

In this article, we demonstrate the experimental implementation of a two-layer
integrated superconducting circuit where two microchips are attached by means of
thermocompression bonding in vacuum. The structures on the surface of the bottom
chip (or \textit{base chip}) and on the underside of the top chip (or
\textit{cap}) are fabricated using standard photolithography techniques. Instead
of a discrete set of indium bump bonds, a continuous thin film of molten indium
serves as bonding medium between the chips. We perform a detailed electrical
characterization of a variety of bonded devices from room temperature
to~\SI{10}{\milli\kelvin} at both dc and microwave frequencies, showing that the
bonding technology can be used in quantum computing applications.

Figure~\ref{Figure1:McRae:Main}~(a) illustrates the geometric characteristics of
a capped device. The base chip consists of an intrinsic silicon substrate of
thickness~\SI{500}{\micro\meter} coated by an aluminum film
with~\SI{150}{\nano\meter} thickness followed by an indium film of equal
thickness. These films are sputtered \textit{in situ} in an ATC-Orion 5 sputter
system from AJA International, Inc. The coplanar waveguide~(CPW) transmission
line visible in Fig.~\ref{Figure1:McRae:Main}~(a) and other on-chip structures
are defined by optical lithography followed by a wet-etch in Transene~A aluminum
etchant.~\footnote{We find that Transene~A works well to etch indium thin films
as well as aluminum.}

The cap consists of a~\SI{350}{\micro\meter} thick silicon wafer with tunnels
trenched by isotropic reactive-ion etching~(RIE) and through holes formed using
a deep~RIE process. After etching, the cap underside is metallized with the same
aluminum-indium process as the base chip.

\begin{figure*}[t!]
	\centering
	\includegraphics[width=0.99\textwidth]{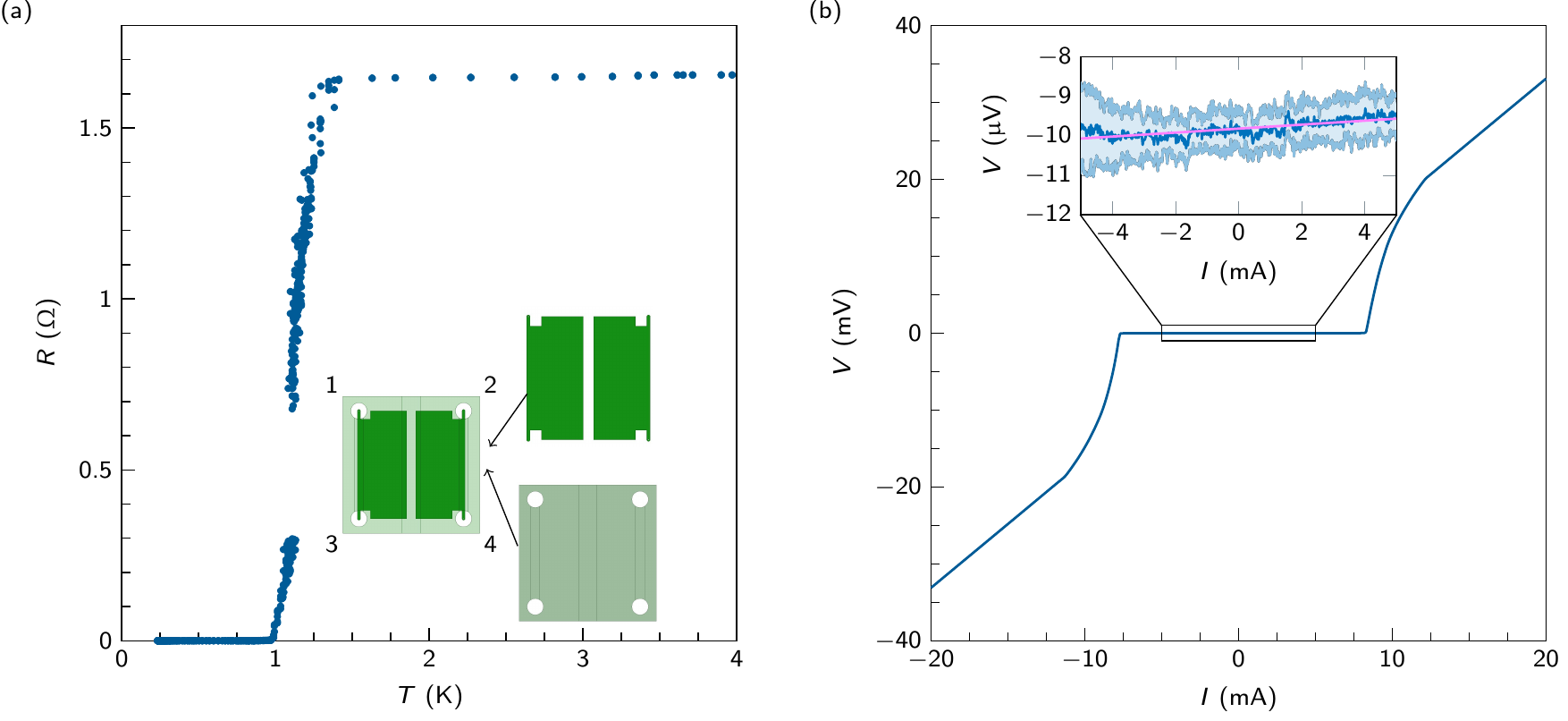}
\caption{Bond characterization at dc. (a) Resistance~$R$ as a function of
temperature~$T$ for the capped device depicted in the inset. Inset: Base chip
with two aluminum-indium islands separated by a dielectric gap. The fully
metallized cap includes a tunnel which, when the chips are bonded, spans the gap
on the base chip. (b) I-V~characteristic curve at~$T \simeq
\SI{10}{\milli\kelvin}$ for the capped device in~(a). Inset: Data and fit
[magenta (middle gray)] below~$I_{\textrm{c}}$.}
	\label{Figure2:McRae:Main}
\end{figure*}

The bonding procedure is realized in a custom-made vacuum chamber with the aid
of an aligning and compressing fixture (see details in the supplementary
material). The base chip and cap are aligned with a horizontal accuracy of less
than~\SI{10}{\micro\meter}, significantly smaller than the device's maximum
allowed tolerances. The chips are subsequently compressed by applying vertical
pressure with the fixture lid. At a system pressure of
approximately~\SI{e-2}{\milli\bar}, the chamber is placed for~\SI{100}{\minute}
on a hot plate at~\SI{190}{\degreeCelsius}, above the indium melting temperature
($\sim\!\!\SI{157}{\degreeCelsius}$). Heating in vacuum prevents the formation
of thick indium oxide~\cite{Schoeller:2006} and results in a strong mechanical
bond without chemical or physical cleaning of the indium films prior to bonding.
In fact, we found that bonded samples can withstand several minutes of
high-power sonication and multiple cooling cycles to~\SI{77}{\kelvin} and
\SI{10}{\milli\kelvin}. Images of a bonded device are shown in
Figs.~\ref{Figure1:McRae:Main}~(b), \ref{Figure1:McRae:Main}(c), and
\ref{Figure1:McRae:Main}(d).

The dc electrical behavior of a bonded device is characterized using the base
chip and cap layouts shown in the inset of
Fig.~\ref{Figure2:McRae:Main}.~\footnote{This design guarantees that a dc
current flows through the indium bond and the metallized tunnel when measuring
the resistance between the two base-chip islands.} By applying a dc current
through ports~$1$ and $4$ and measuring the voltage across ports~$2$ and $3$, we
find a room temperature dc resistance~$R \simeq \SI{2.785}{\ohm}$. This
resistance is due to the room temperature resistance of the aluminum-indium
films on both the base chip and cap as well as the bond resistance between the
two chips. We realize detailed numerical simulations of the device under
test~(DUT) by means of ANSYS Q3D
Extractor~\footnote{\url{http://www.ansys.com/Products/Electronics/ANSYS-Q3D-Extractor}}
 and find a theoretical~$R \simeq \SI{1.323}{\ohm}$. The discrepancy between 
the measured and simulated resistance is likely due to the bond resistance (not 
included in the Q3D model), bond inhomogeneity, or both; possible contributors 
to these effects are the presence of native indium oxide before bonding, 
inter-diffusion of aluminum and indium, or a tilt between base chip and cap.

Bond inhomogeneity can be understood by modeling the five metallic layers of a
bonded device \textendash~aluminum, indium, bond region, indium, and aluminum
\textendash~as a large set of flux tubes directed from the base chip to the cap.
At room temperature, each tube has a resistance~$R_{\textrm{tube}}$; the total
resistance~$R$ is approximately the parallel resistance of all flux tubes. Flux
tubes in an unbonded region are open circuits with resistance~$R_{\textrm{tube}}
\sim \infty$ and cause~$R$ to increase. Following this model, the ratio between
measured and simulated resistance indicates that approximately~\SI{50}{\percent}
of the DUT is bonded. This result is verified by breaking apart bonded devices
and inspecting optically the surfaces of the bond region (see supplementary
material).

Figure~\ref{Figure2:McRae:Main}~(a) shows a four-point measurement of~$R$ as a
function of temperature~$T$ for the bonded device shown in the inset. Below the
superconducting transition temperature of aluminum, $T \simeq
\SI{1.2}{\kelvin}$, $R$ is approximately the bond resistance. The data points in
the figure are obtained by measuring the device's current-voltage~(I-V)
characteristic curves at various temperatures and fitting their slope.
Figure~\ref{Figure2:McRae:Main}~(b) shows the I-V~curve measured at~$T \simeq
\SI{10}{\milli\kelvin}$. We find a critical current intensity for the
aluminum-indium films, $I_{\textrm{c}} \simeq \SI{7}{\milli\ampere}$. An
ensemble of measurements well below~$I_{\textrm{c}}$ is reported in the inset of
Fig.~\ref{Figure2:McRae:Main}~(b). From the least-squares best fit, we obtain a
bond resistance~$R \simeq 50 \mp \SI{2}{\micro\ohm}$, which is less than
approximately~\SI{515}{\nano\ohm\per\milli\meter\squared} assuming a
\SI{50}{\percent} bond area. This resistance is likely due to the native indium
oxide film initially present on the chips, preventing the bond region from
becoming superconductive at low temperatures.

\begin{figure*}[t!]
	\centering
	\includegraphics[width=0.99\textwidth]{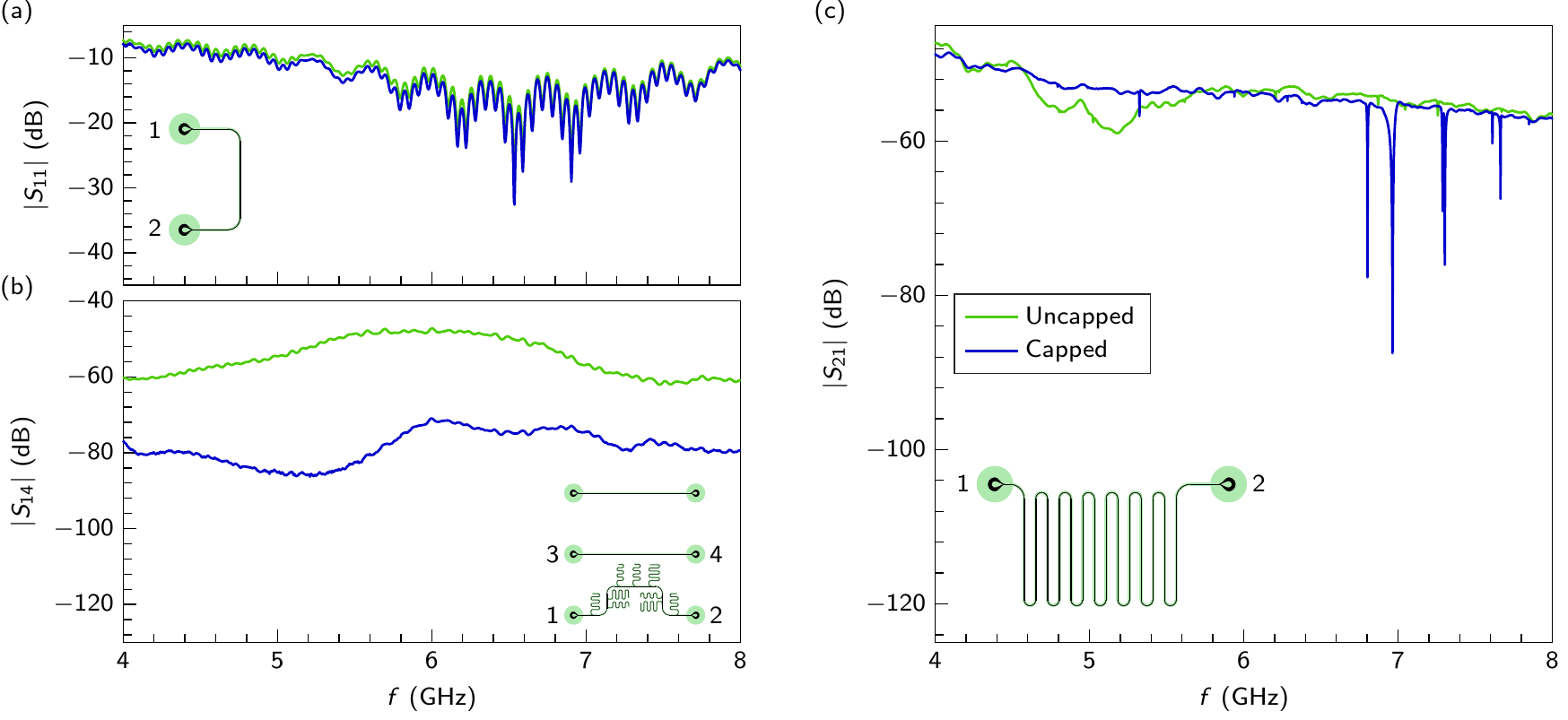}
\caption{Characterization at microwave frequencies of the uncapped and capped
CPW transmission lines shown in insets; black lines refer to structures on the
base chip and light green (light gray) shades indicate metallized tunnels and
through holes in the cap. Data for uncapped devices is plotted in light green
(light gray) and for capped devices in dark blue (dark gray). (a) Magnitude of
the room temperature reflection coefficient at port~$1$, $\abs{S_{11}}$, as a
function of frequency~$f$. (b) Magnitude of the room temperature isolation
coefficient between ports~$1$ and $4$, $\abs{S_{14}}$, vs.~$f$. (c) Magnitude of
the transmission coefficient~$\abs{S_{21}}$ measured at~\SI{10}{\milli\kelvin}.}
	\label{Figure3:McRae:Main}
\end{figure*}

Figure~\ref{Figure3:McRae:Main} displays the microwave characterization of three
capped and uncapped devices with layouts shown in the insets. The measurements
are realized by means of a vector network analyzer~(VNA) from Keysight
Technologies Inc., model~PNA-X N5242A; details on the measurement setups can be
found in B{\'{e}}janin~\textit{et al.}~\cite{Bejanin:2016}

The measurements in Fig.~\ref{Figure3:McRae:Main}~(a) demonstrate that the
bonding process and the addition of the cap \textendash~including the tunnel
mouth \textendash~do not noticeably increase reflections, which are instead
dominated by the three-dimensional wires.~\cite{Bejanin:2016}

Figure~\ref{Figure3:McRae:Main}~(b) shows a measurement of the isolation
coefficient between two adjacent transmission lines (see inset). At microwave
frequencies a signal injected at port~$1$ or $2$ can leak to ports~$3$ and $4$,
generating signal crosstalk. Due to the three-dimensional wires, signal
crosstalk for the uncapped device is already very low.~\cite{Bejanin:2016} The
cap further reduces crosstalk by more than~\SI{10}{\deci\bel} across the entire
measurement bandwidth. These results may have important implications to quantum
computing, where crosstalk has been identified as a major source of
error.~\cite{Najafi-Yazdi:2017}

Reflection and isolation measurements are performed at room temperature in order
to maintain the DUT reference planes as close as possible to the VNA ports.
However, the line shown in the inset of Fig.~\ref{Figure3:McRae:Main}~(c) is
characterized by a room-temperature loss sufficiently large to mask any
abnormalities in transmission measurements at microwave frequencies. We
therefore measure this line at~$\sim\!\!\SI{10}{\milli\kelvin}$, where both the
indium and aluminum films are in the superconducting state.

Figure~\ref{Figure3:McRae:Main}~(c) shows clean transmission for both uncapped
and capped devices up to~$f \simeq \SI{6.8}{\giga\hertz}$. At higher
frequencies, we observe a series of pronounced resonances in the transmission
coefficient of the capped device. The simulations in Fig.~S3 of the
supplementary material and the results in Fig.~\ref{Figure3:McRae:Main}~(a)
indicate that these resonances are not due to the presence of the tunnel. We
believe they are caused by bond inhomogeneity, shown in Fig.~S2 of the
supplementary material, resulting in unwanted resonances similar to the slotline
modes observed by Wenner~\textit{et al.}~\cite{Wenner:2011:a}

As a proof of concept for quantum computing applications, we fabricate and
measure a set of capped superconducting CPW resonators. The device layout is
sketched in the inset of Fig.~\ref{Figure3:McRae:Main}~(b), which shows
nine~$\lambda / 4 $-wave resonators coupled to line~$1-2$ in a multiplexed
design.~\cite{Bejanin:2016}

The resonators are characterized by measuring their internal quality factor
at~$\sim\!\!\SI{10}{\milli\kelvin}$ and for a mean photon occupation
number~$\langle n_{\textrm{ph}} \rangle \simeq 1$, similar to the excitation
power used in quantum computing operations. The main resonator parameters are
reported in Table~\ref{Table1:McRae:Main}.~\footnote{Five of the nine capped
resonators were damaged during fabrication and, thus, their parameters are not
reported in Table~\ref{Table1:McRae:Main}. The fits for the last two capped
resonators in the table are poor; the corresponding quality factors are not
shown.} The measured data and fits for the second pair of resonators in
Table~\ref{Table1:McRae:Main} are shown in Fig.~S4 of the supplementary
material. Note that the resonance frequency of an uncapped resonator shifts with
the addition of the cap.

In the supplementary material, we determine that the internal quality factor of
a capped resonator is approximately~\SI{1}{\percent} larger than that of an
uncapped resonator due to the vacuum participation. However, this effect is
significantly smaller than the quality factor fluctuations over
time~\footnote{By measuring a resonator over a time period of ten hours, we
found a standard deviation as large as half of the mean quality
factor.~\cite{Neill:2013}} and, thus, it cannot be resolved in our measurements.
In fact, accounting for time fluctuations, the quality factors for capped and
uncapped resonators in Table~\ref{Table1:McRae:Main} are approximately equal.

\begin{table}[b!]
\caption{Resonance frequencies and internal quality factors $f^{\textrm{uc}}_0$,
$f^{\textrm{c}}_0$ and $Q^{\textrm{uc}}_{\textrm{i}}$,
$Q^{\textrm{c}}_{\textrm{i}}$, respectively, for a set of four uncapped and
capped resonators.}
\begin{center}
	\begin{ruledtabular}
		\begin{tabular}{ccccccc}
			\raisebox{0mm}[3mm][0mm]{$f^{\textrm{uc}}_0$ 
			\footnotesize{(\SI{}{\giga\hertz})}} & 
			$Q^{\textrm{uc}}_{\textrm{i}}$ & & $f^{\textrm{c}}_0$ 
			\footnotesize{(\SI{}{\giga\hertz})} & $Q^{\textrm{c}}_{\textrm{i}}$ 
			\\
			\hline
			\hline
			\raisebox{0mm}[3mm][0mm]{$4.252$} & $37700$ & & $4.033$ & $20230$ \\
			\raisebox{0mm}[3mm][0mm]{$4.448$} & $52100$ & & $4.982$ & $22510$ \\
			\raisebox{0mm}[3mm][0mm]{$4.722$} & $41830$ & & $\cdots$ & $\cdots$ 
			\\
			\raisebox{0mm}[3mm][0mm]{$4.913$} & $44840$ & & $\cdots$ & $\cdots$ 
			\\
			\vspace{-4.5mm}
		\end{tabular}
	\end{ruledtabular}
\end{center}
	\label{Table1:McRae:Main}
	\vspace{-3.5mm}
\end{table}

In conclusion, we have developed and characterized a hot thermocompression
bonding technology in vacuum using indium thin films as bonding agent. Our
results show that this technology can be readily used to implement an integrated
multilayer architecture, combining the fabrication advantages of two-dimensional
superconducting qubits~\cite{Clarke:2008} and the long coherence of
micromachined three-dimensional cavities.~\cite{Brecht:2015} This bonding
technology is compatible with the quantum socket design, paving the way toward
the implementation of extensible quantum computing architectures as proposed by
B{\'{e}}janin~\textit{et al.}~\cite{Bejanin:2016}

In future implementations, we will improve the bonding procedure by including a
cleaning step with an acid buff to remove native indium oxide prior to bonding.
Additionally, we will use slightly thicker indium films
($\sim\!\!\SI{1}{\micro\meter}$), a lower bonding pressure
($\sim\!\!\SI{e-6}{\milli\bar}$), \textit{in vacuo} chip alignment and
compression, as well as a higher and more homogeneous compression by means of an
hydraulic press. Finally, we will add an inter-diffusion barrier between the
aluminum and indium films.

\vspace{2.5mm}
See supplementary materials for details on the thermocompression bonding setup,
bond inhomogeneity, transmission simulations, resonator data and fits, and
vacuum participation.

\vspace{2.5mm}
We acknowledge the Natural Sciences and Engineering Research Council of Canada
and the Canadian Microelectronics Corporation Microsystems. We thank the Quantum
NanoFab Facility at the University of Waterloo as well as the Toronto
Nanofabrication Centre for their support with device fabrication, as well as
F.~Deppe and J.Y.~Mutus for fruitful discussions.

\clearpage

\pagebreak

\begin{center}
\textbf{\large Supplementary Materials for ``Thermocompression Bonding Technology for
Multilayer Superconducting Quantum Circuits''}
\end{center}

\newcommand{\beginsupplement}{%
	\setcounter{section}{0}
	\renewcommand{\thesection}{S\arabic{section}}%
	\setcounter{subsection}{0}
	\renewcommand{\thesubsection}{S\Roman{subsection}}%
	\setcounter{subsubsection}{0}
	\renewcommand{\thesubsubsection}{S\Alph{subsubsection}}%
	\titleformat{\subsubsection}[block]{\bfseries\centering}{\thesubsubsection.}{1em}{}
	\setcounter{table}{0}
	\renewcommand{\thetable}{S\arabic{table}}%
	\setcounter{figure}{0}
	\renewcommand{\thefigure}{S\arabic{figure}}%
	\setcounter{equation}{0}
	\renewcommand{\theequation}{S\arabic{equation}}%
	}

\setcounter{page}{1}
\makeatletter

\renewcommand{\bibnumfmt}[1]{[S#1]}
\renewcommand{\citenumfont}[1]{S#1}

In this supplementary materials, we provide further details on the
thermocompression bonding setup. We characterize bond inhomogeneity with optical
imaging. We present numerical simulations of the transmission coefficient at
microwave frequencies for various configurations of capped and uncapped coplanar
waveguide~(CPW) transmission lines. We show the measured data and fits for a
capped and uncapped superconducting resonator at~\SI{10}{\milli\kelvin}.
Finally, we calculate the ratio of the internal quality factor for a capped and
uncapped resonator, thus estimating the effect of vacuum participation.

\section*{S1: THERMOCOMPRESSION BONDING SETUP AND PROCEDURE}
	\label{THERMOCOMPRESSION:BONDING:SETUP:AND:PROCEDURE}

\begin{figure*}[t!]
	\centering
	\includegraphics[width=0.99\textwidth]{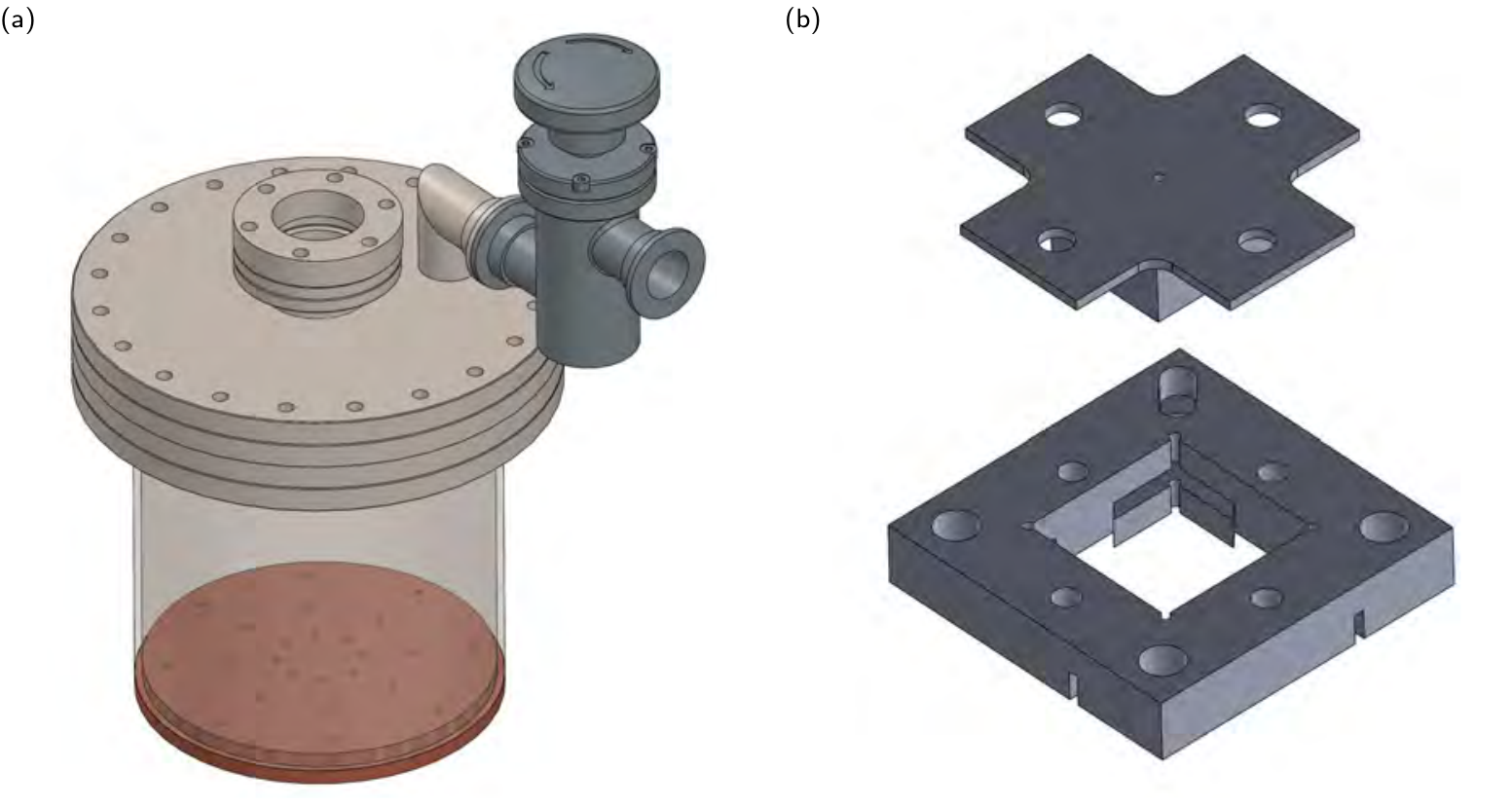}
\caption{Thermocompression bonding setup. (a) Vacuum chamber. (b)
Aligning-compressing fixture comprising a square washer, an adjustable edge
corner, and a lid. The channels used to evacuate the interior of the washer when
tightened to the chamber's bottom are visible.}
\end{figure*}

Figure~S1 shows a computer-aided design of the custom-made vacuum chamber and
aligning-compressing fixture used for the thermocompression bonding. The bottom
of the chamber is made from a~\SI{10}{\milli\meter} thick copper plate, ensuring
a high thermal conductivity and heat capacity. The bottom surface of the plate
is mirror polished each time before placing the chamber on the hot plate. The
top surface features a set of threaded screw holes, where aligning-compressing
fixtures with different dimensions can be fixed; the chamber is designed to
process up to~$3$-in. wafers. The chamber's wall is a~\SI{30}{\centi\meter}
high, \SI{3}{\milli\meter} thick hollow cylinder made from stainless steel, the
low thermal conductivity of which ensures little or no heating of the top part
of the chamber. The top edge of the cylindrical wall is welded to a
$6$-in.~ConFlat~(CF) flange made from 304L stainless steel.~\footnote{A new
chamber is being designed made from 316LN stainless steel with low magnetic
permeability~$\mu \leq 1.005$, reducing magnetic contamination of the samples.}
The CF~flange features a knife-edge seal mechanism; the sealing element is a
fully annealed copper gasket. The gasket is made from $1/4$ hard, high purity,
oxygen-free~(OFHC) copper, which allows for system pressures as low
as~\SI{e-12}{\milli\bar} at operating temperatures up
to~\SI{450}{\degreeCelsius}. A ultra-high vacuum, temperature-resistant valve is
used to connect the CF~flange to a pump. A CF~flanged Kodial glass viewport
(sealed with a silver plated copper gasket) permits the observation of the
chamber's interior during processing.

The aligning-compressing fixture is a square washer with inner
dimensions~$\SI{16.5}{\milli\meter} \times \SI{16.5}{\milli\meter}$ made from
stainless steel and featuring an adjustable edge corner. This corner can be
moved along a groove at the bottom of the washer, allowing alignment between the
base chip and cap. In this work, we use square chips with
dimensions~$\SI{15}{\milli\meter} \times \SI{15}{\milli\meter}$. Note that the
base chip is in direct contact with the copper plate, guaranteeing good
thermalization. The aligning procedure comprises three steps: First, the base
chip is placed inside the washer at the bottom of the chamber and pushed against
the washer's corner opposite to the adjustable corner; second, the cap is
manually dropped on the base chip, pre-aligning the chips as accurately as
possible; third, the two chips are aligned with the adjustable corner, which is
finally fixed to the chamber with a screw. The manual pre-alignment in step two
is sufficient to prevent damage of the on-chip structures due to relative
dragging of the chips during step three. The lid shown in Fig.~S1~(b) is used to
apply pressure on the aligned chips by means of four screws. We find that little
pressure is required to obtain a mechanically strong bond.

After closing the chamber, we evacuate it with a molecular pump from Pfeiffer
Vacuum GmbH for~\SI{30}{\minute} to reach a system pressure
of~\SI{1.5e-2}{\milli\bar}. While pumping we monitor the chamber leak rate,
which is typically on the order of~\SI{e-10}{\milli\bar\liter\per\second}. Once
the chamber is placed on the hot plate, the temperature initially fluctuates
reaching a steady state in approximately~\SI{10}{\minute}; at that time we start
counting~\SI{100}{\minute}. Pumping is continued during the entire heating
process, as well as during a cooling period when the chamber is placed to rest
on a thick aluminum plate.

Our bonding procedure is simple and highly reproducible, with a yield
of~\SI{80}{\percent} for a total of~$15$ bonded samples.

When performing thermocompression bonding of devices with superconducting
qubits, the Josephson tunnel junctions that make the qubits will be heated at
high temperature. In order to verify whether this process damages the Josephson
junctions, we fabricate a set of aluminum-aluminum oxide-aluminum junctions
using e-beam lithography and evaporation. The junctions' area
is~$\SI{350}{\nano\meter} \times \SI{350}{\nano\meter}$ and their room
temperature resistance~$\sim\!\!\SI{5}{\kilo\ohm}$. We perform a preliminary
test by heating the junctions to~\SI{190}{\degreeCelsius} for~\SI{5}{\minute} at
atmospheric pressure, finding that most of the junctions survive the process. In
addition, we find that the post-heat room temperature resistance decreases to
approximately~\SI{4}{\kilo\ohm}. A complete study of heating effects on
submicron sized Josephson junctions was performed by Koppinen \textit{et
al.}~\cite{Koppinen:2006} By heating the junctions up
to~\SI{500}{\degreeCelsius} in high vacuum (less than~\SI{e-6}{\milli\bar}), the
authors found a stabilization of the junctions' tunneling resistance as well as
a reduction of the junctions' rate of aging. These results are very encouraging,
indicating that the hot bonding process may even improve the quality of the
Josephson junctions.

\section*{S2: BOND INHOMOGENEITY}
	\label{BOND:INHOMOGENEITY}

Bond inhomogeneity can be characterized by breaking apart a bonded device and
imaging the base chip and cap bond surfaces optically, as shown in Fig.~S2. The
images refer to the device outlined in the inset of Fig.~3~(c) in the main text
and are taken by means of a handheld digital microscope.

The indium film within the boundary of the through holes on the base chip [see
Fig.~S2~(a)] is heated during the bonding process, but not bonded to the cap. We
can thus use the color of the indium film in this region as a reference to
discern bonded from not bonded regions. We determine that the region near the
center of the base chip and cap was bonded well, whereas the area around the
edges of the two chips was not bonded. In this case,
approximately~\SI{50}{\percent} of the chips' area was bonded well. We find
similar results in other devices. This effect is likely due to the compressing
fixture lid that was designed to be smaller than the chips' area, thus applying
pressure only to the center of the chips. In the conclusions of the main text we
propose a remedy to this issue.

\begin{figure*}[t!]
	\centering
	\includegraphics[width=0.99\textwidth]{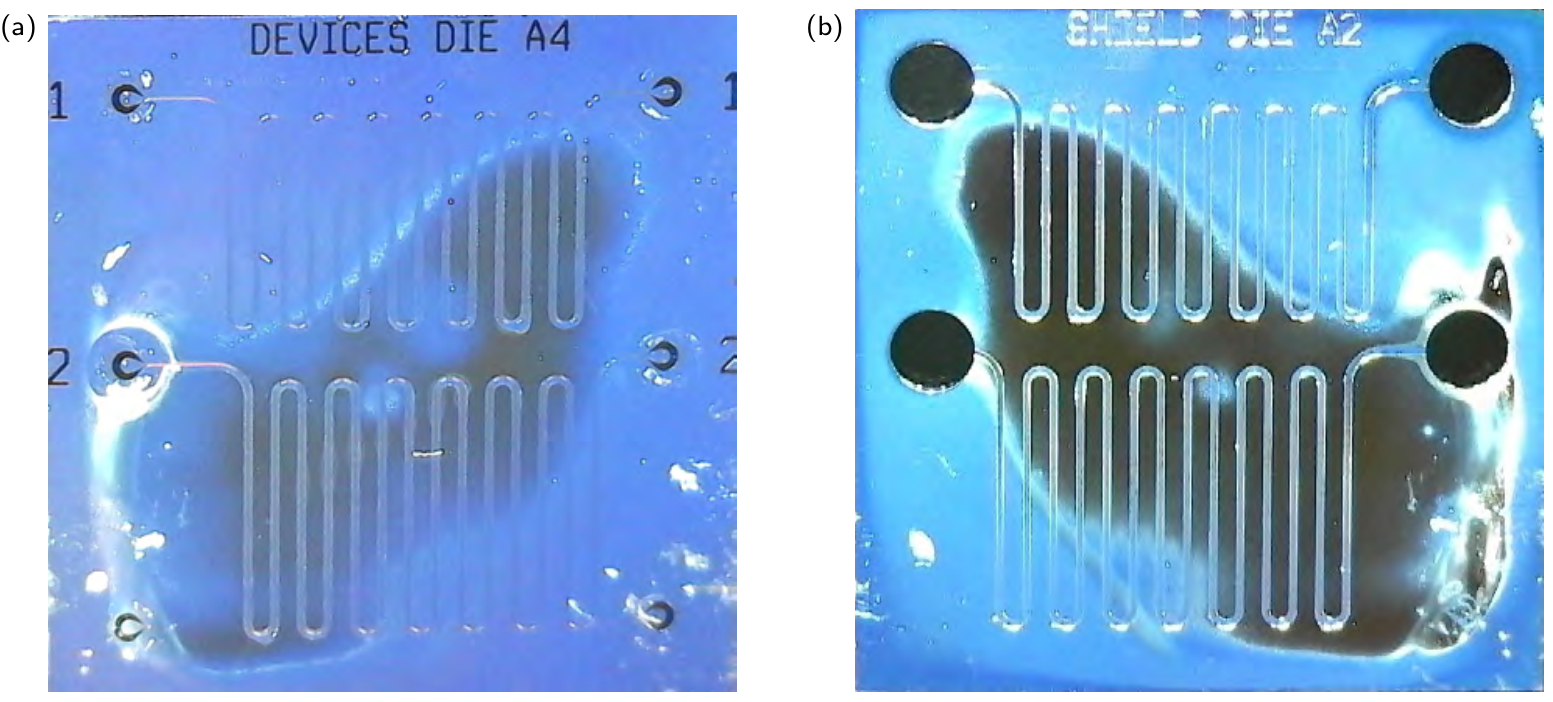}
\caption{Optical characterization of bond inhomogeneity. (a) Image of the base
chip after bonding. The marks left by the three-dimensional wires on trace~$1$
and $2$ are clearly visible. (b) Image of the cap after bonding.}
\end{figure*}

\section*{S3: TUNNEL MOUTH MICROWAVE SIMULATIONS}
	\label{TUNNEL:MOUTH:MICROWAVE:SIMULATIONS}

The edge of a through hole where the cap tunnel begins (the \textit{tunnel
mouth}) represents a boundary condition for the electromagnetic field associated
with a capped transmission line. The line shown in the inset of Fig.~3~(c) in
the main text is characterized by two of such boundary conditions. In order to
determine whether these conditions generate unwanted resonance modes, we
simulate the transmission coefficient~$S_{21}$ for this line and compare it to
that of an uncapped line and of a capped line without tunnel mouths (i.e.,
covered by an infinitely long tunnel). The numerical simulations are performed
with ANSYS
HFSS,~\footnote{\url{http://www.ansys.com/products/electronics/ansys-hfss}}
assuming perfect conductors and lossless CPW transmission lines with equal
geometric characteristics.

The graphs displayed in Fig.~S3 reveal almost perfect transmission for the three
simulated configurations, with less than~\SI{0.1}{\decibel} of loss due to
slight impedance mismatch. We can safely conclude that the unwanted resonances
shown in Fig.~3~(c) of the main text are not due to the presence of the tunnel
mouths.

\begin{figure}[b!]
	\centering
	\includegraphics[width=0.495\textwidth]{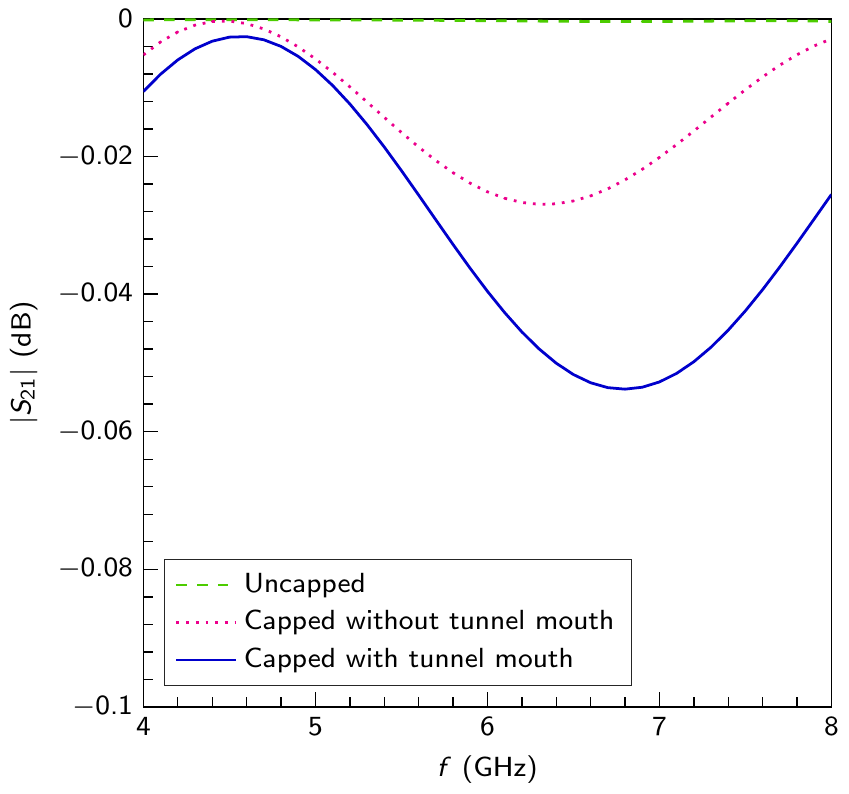}
\caption{Magnitude of the simulated transmission coefficient~$\abs{S_{21}}$ as
function of frequency~$f$ for an uncapped, capped without tunnel mouth, and
capped with tunnel mouth CPW transmission line. The chosen frequency range is
the same as in Fig.~3~(c) of the main text.}
\end{figure}

\section*{S4: CAPPED SUPERCONDUCTING CPW RESONATORS}
	\label{CAPPED:SUPERCONDUCTING:CPW:RESONATORS}

Figure~S4 shows data and fits for the second pair of uncapped and capped
superconducting resonators reported in Table~1 of the main text. The displayed
data for the capped resonator is the ensemble average of~$2$ measured traces,
whereas only one trace is measured for the uncapped resonator. In both cases,
each data point is obtained setting the vector network analyzer~(VNA) to an
intermediate frequency bandwidth~$\Delta f_{\textrm{IF}} = \SI{1}{\hertz}$.

\begin{figure*}[t!]
	\centering
	\includegraphics[width=0.99\textwidth]{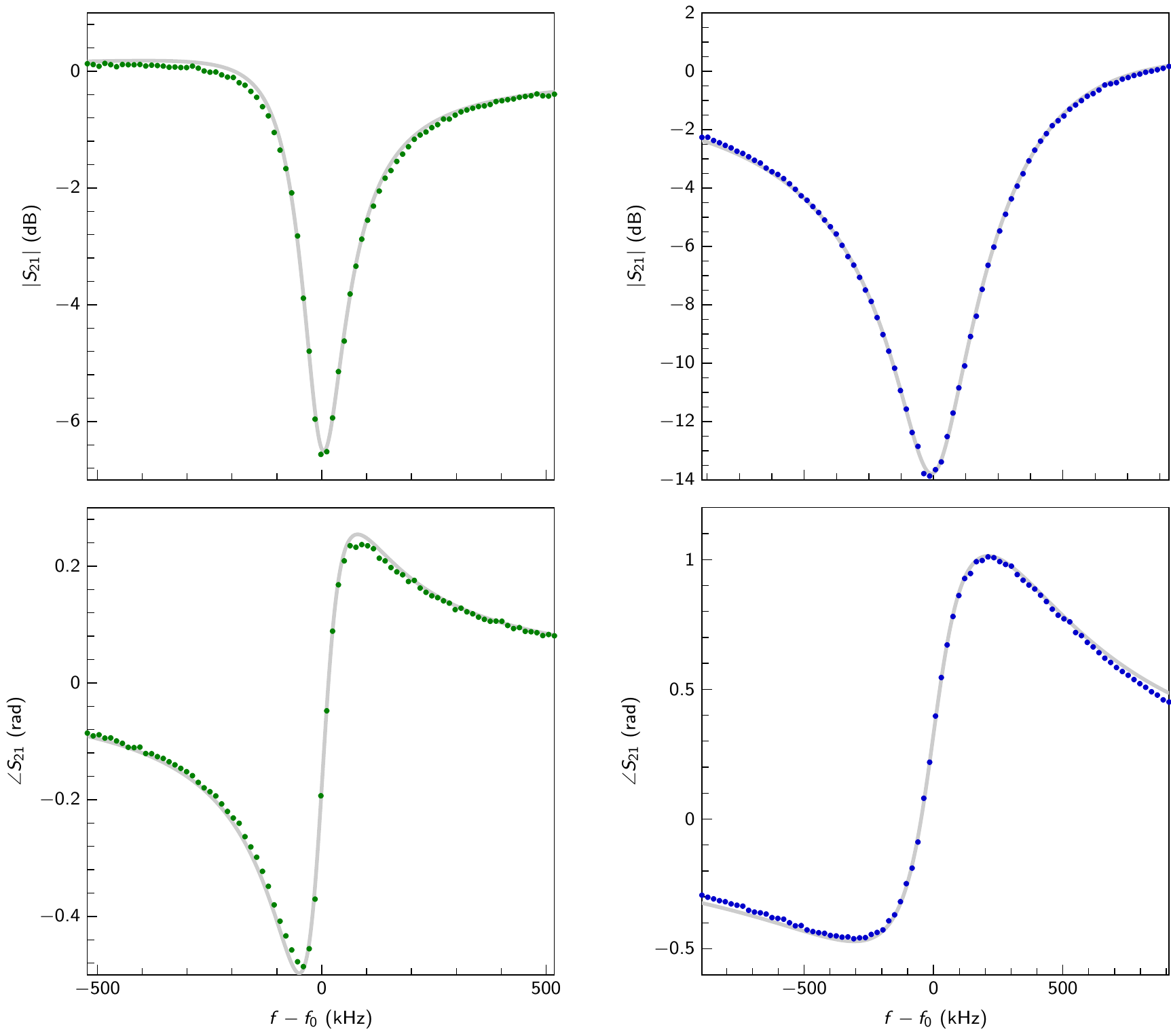}
\caption{Uncapped (left panels) and capped (right panels) resonator measurements
(dots) at low power [i.e., $\langle n_{\textrm{ph}} \rangle \simeq 1$ (cf.~main
text)]. The resonator transmission magnitude~$\abs{S_{21}}$ (above) and phase
angle~$\angle S_{21}$ (below) are fitted as in B{\'{e}}janin~\textit{et
al.}~\cite{Bejanin:2016} (light gray).}
\end{figure*}

\section*{S5: VACUUM CONTRIBUTION TO CAPPED RESONATORS QUALITY FACTOR}
	\label{VACUUM:CONTRIBUTION:TO:CAPPED:RESONATORS:QUALITY:FACTOR}

In this section, we estimate the effect of a metallized cap on the internal
quality factor~$Q_{\textrm{i}}$ of a superconducting CPW transmission line
resonator. The addition of a grounded cap above a CPW resonator forces some of
the electric field lines to be distributed from the base chip to the cap, away
from the base chip substrate. This increases the contribution of vacuum to the
mode volume of a capped resonator compared to the case of an uncapped resonator.
Assuming all metallic structures to be perfect conductors, the internal quality
factor is solely due to dielectric losses and, thus, it can be found by
inverting the loss tangent as,~\cite{Simons:2001}
\begin{equation}
Q_{\textrm{i}} = \dfrac{\varepsilon'_{\textrm{e}}}{\varepsilon''_{\textrm{e}}} 
\quad ,
	\label{Equation:01S}
\end{equation}
where~$\varepsilon_{\textrm{e}} = \varepsilon'_{\textrm{e}} - j
\varepsilon''_{\textrm{e}}$ is the effective electric complex permittivity of
the CPW transmission line with real and imaginary
parts~$\varepsilon'_{\textrm{e}}$ and $\varepsilon''_{\textrm{e}}$, respectively
($j^2 = -1$).

The effective electric permittivity of a capped CPW transmission line can be
calculated using Eq.~(2.39) in Simons~\cite{Simons:2001}
\begin{equation}
\varepsilon^{\textrm{c}}_{\textrm{e}} = 1 + q_3 \left( 
\varepsilon_{\textrm{r}1} - 1 \right) 
\quad ,
	\label{Equation:02S}
\end{equation}
where~$q_3$ is the partial filling factor dependent on the device geometry [see
Eq.~(2.40) in Simons~\cite{Simons:2001}] and $\varepsilon_{\textrm{r}1} =
\varepsilon'_{\textrm{r}1} - j \varepsilon''_{\textrm{r}1}$ is the relative
electric complex permittivity of the base chip substrate (in our case silicon)
with thickness~$h_1$. Hereafter, we assume~$h_1 \to \infty$ (a reasonable
approximation as the silicon substrates are~\SI{500}{\micro\meter} thick, much
thicker than any of the other structures; see main text for all relevant
dimensions). Note that Eq.~(\ref{Equation:02S}) is applicable as the tunnel
sidewalls are much farther away from the conductor than the in-plane ground
planes, $T \gg H$ (see Fig.~1 in the main text).

Inserting Eq.~(\ref{Equation:02S}) into Eq.~(\ref{Equation:01S}), we obtain the
capped internal quality factor
\begin{equation}
Q^{\textrm{c}}_{\textrm{i}} = \dfrac{1 + q_3 \left( \varepsilon'_{\textrm{r}1} 
- 1 \right)}{q_3 \varepsilon''_{\textrm{r}1}} \quad .
	\label{Equation:03S}
\end{equation}
\vspace{0mm}

In the case of an uncapped CPW transmission line, the effective electric
permittivity is given by~\cite{Simons:2001}
\begin{equation}
\varepsilon^{\textrm{uc}}_{\textrm{e}} = \dfrac{1 + 
\varepsilon_{\textrm{r}1}}{2}
	\label{Equation:04S}
\end{equation}
and the internal quality factor is given by
\begin{equation}
Q^{\textrm{uc}}_{\textrm{i}} = \dfrac{1 + 
\varepsilon'_{\textrm{r}1}}{\varepsilon''_{\textrm{r}1}} \quad .
	\label{Equation:05S}
\end{equation}

The ratio between the uncapped and capped internal quality factors is thus
\begin{equation}
\dfrac{Q^{\textrm{uc}}_{\textrm{i}}}{Q^{\textrm{c}}_{\textrm{i}}} = 
\dfrac{\left( 1 + \varepsilon'_{\textrm{r}1} \right) q_3}{\left( 
\varepsilon'_{\textrm{r}1} - 1 \right) q_3 + 1} \quad .
	\label{Equation:06S}
\end{equation}
Using the dimensions~$S$, $W$, and $h_4 = H$ reported in the main text and
assuming~$\varepsilon'_{\textrm{r}1} = 11$ for silicon, we find~$q_3 \simeq
0.4722$ and, thus, $Q^{\textrm{uc}}_{\textrm{i}} / Q^{\textrm{c}}_{\textrm{i}}
\simeq 0.99$. As a consequence, the increase in vacuum participation due to the
addition of the cap increases the internal quality factor by
approximately~\SI{1}{\percent}. This is a very small effect for the devices
presented in this work, where other loss mechanisms such as the presence of
native indium oxide on both the base chip and cap, the low quality thin-film
sputter deposition, and possibly the bonding procedure itself outweigh the
benefits of a higher vacuum participation. However, a careful design and a
suitable fabrication and cleaning process may be used to take advantage of this
effect to make capped devices (e.g., qubits) with lower error rates than similar
uncapped devices.


\begin{thebibliography}{33}%
\makeatletter
\providecommand \@ifxundefined [1]{%
 \@ifx{#1\undefined}
}%
\providecommand \@ifnum [1]{%
 \ifnum #1\expandafter \@firstoftwo
 \else \expandafter \@secondoftwo
 \fi
}%
\providecommand \@ifx [1]{%
 \ifx #1\expandafter \@firstoftwo
 \else \expandafter \@secondoftwo
 \fi
}%
\providecommand \natexlab [1]{#1}%
\providecommand \enquote  [1]{``#1''}%
\providecommand \bibnamefont  [1]{#1}%
\providecommand \bibfnamefont [1]{#1}%
\providecommand \citenamefont [1]{#1}%
\providecommand \href@noop [0]{\@secondoftwo}%
\providecommand \href [0]{\begingroup \@sanitize@url \@href}%
\providecommand \@href[1]{\@@startlink{#1}\@@href}%
\providecommand \@@href[1]{\endgroup#1\@@endlink}%
\providecommand \@sanitize@url [0]{\catcode `\\12\catcode `\$12\catcode
  `\&12\catcode `\#12\catcode `\^12\catcode `\_12\catcode `\%12\relax}%
\providecommand \@@startlink[1]{}%
\providecommand \@@endlink[0]{}%
\providecommand \url  [0]{\begingroup\@sanitize@url \@url }%
\providecommand \@url [1]{\endgroup\@href {#1}{\urlprefix }}%
\providecommand \urlprefix  [0]{URL }%
\providecommand \Eprint [0]{\href }%
\providecommand \doibase [0]{http://dx.doi.org/}%
\providecommand \selectlanguage [0]{\@gobble}%
\providecommand \bibinfo  [0]{\@secondoftwo}%
\providecommand \bibfield  [0]{\@secondoftwo}%
\providecommand \translation [1]{[#1]}%
\providecommand \BibitemOpen [0]{}%
\providecommand \bibitemStop [0]{}%
\providecommand \bibitemNoStop [0]{.\EOS\space}%
\providecommand \EOS [0]{\spacefactor3000\relax}%
\providecommand \BibitemShut  [1]{\csname bibitem#1\endcsname}%
\let\auto@bib@innerbib\@empty
\bibitem [{\citenamefont {Ladd}\ \emph {et~al.}(2010)\citenamefont {Ladd},
  \citenamefont {Jelezko}, \citenamefont {Laflamme}, \citenamefont {Nakamura},
  \citenamefont {Monroe},\ and\ \citenamefont {O'Brien}}]{Ladd:2010}%
  \BibitemOpen
  \bibfield  {author} {\bibinfo {author} {\bibfnamefont {T.~D.}\ \bibnamefont
  {Ladd}}, \bibinfo {author} {\bibfnamefont {F.}~\bibnamefont {Jelezko}},
  \bibinfo {author} {\bibfnamefont {R.}~\bibnamefont {Laflamme}}, \bibinfo
  {author} {\bibfnamefont {Y.}~\bibnamefont {Nakamura}}, \bibinfo {author}
  {\bibfnamefont {C.}~\bibnamefont {Monroe}}, \ and\ \bibinfo {author}
  {\bibfnamefont {J.~L.}\ \bibnamefont {O'Brien}},\ }\bibfield  {title}
  {\enquote {\bibinfo {title} {Quantum computers},}\ }\href {\doibase
  10.1038/nature08812} {\bibfield  {journal} {\bibinfo  {journal} {Nature}\
  }\textbf {\bibinfo {volume} {464}},\ \bibinfo {pages} {45--53} (\bibinfo
  {year} {2010})}\BibitemShut {NoStop}%
\bibitem [{\citenamefont {Monz}\ \emph {et~al.}(2011)\citenamefont {Monz},
  \citenamefont {Schindler}, \citenamefont {Barreiro}, \citenamefont {Chwalla},
  \citenamefont {Nigg}, \citenamefont {Coish}, \citenamefont {Harlander},
  \citenamefont {H{\"a}nsel}, \citenamefont {Hennrich},\ and\ \citenamefont
  {Blatt}}]{Monz:2011}%
  \BibitemOpen
  \bibfield  {author} {\bibinfo {author} {\bibfnamefont {T.}~\bibnamefont
  {Monz}}, \bibinfo {author} {\bibfnamefont {P.}~\bibnamefont {Schindler}},
  \bibinfo {author} {\bibfnamefont {J.~T.}\ \bibnamefont {Barreiro}}, \bibinfo
  {author} {\bibfnamefont {M.}~\bibnamefont {Chwalla}}, \bibinfo {author}
  {\bibfnamefont {D.}~\bibnamefont {Nigg}}, \bibinfo {author} {\bibfnamefont
  {W.~A.}\ \bibnamefont {Coish}}, \bibinfo {author} {\bibfnamefont
  {M.}~\bibnamefont {Harlander}}, \bibinfo {author} {\bibfnamefont
  {W.}~\bibnamefont {H{\"a}nsel}}, \bibinfo {author} {\bibfnamefont
  {M.}~\bibnamefont {Hennrich}}, \ and\ \bibinfo {author} {\bibfnamefont
  {R.}~\bibnamefont {Blatt}},\ }\bibfield  {title} {\enquote {\bibinfo {title}
  {14-qubit entanglement: Creation and coherence},}\ }\href {\doibase
  https://doi.org/10.1103/PhysRevLett.106.130506} {\bibfield  {journal}
  {\bibinfo  {journal} {Physical Review Letters}\ }\textbf {\bibinfo {volume}
  {106}},\ \bibinfo {pages} {130506} (\bibinfo {year} {2011})}\BibitemShut
  {NoStop}%
\bibitem [{\citenamefont {Kelly}\ \emph {et~al.}(2015)\citenamefont {Kelly},
  \citenamefont {Barends}, \citenamefont {Fowler}, \citenamefont {Megrant},
  \citenamefont {Jeffrey}, \citenamefont {White}, \citenamefont {Sank},
  \citenamefont {Mutus}, \citenamefont {Campbell}, \citenamefont {Chen},
  \citenamefont {Chen}, \citenamefont {Chiaro}, \citenamefont {Dunsworth},
  \citenamefont {Hoi}, \citenamefont {Neill}, \citenamefont {O'Malley},
  \citenamefont {Quintana}, \citenamefont {Roushan}, \citenamefont
  {Vainsencher}, \citenamefont {Wenner}, \citenamefont {Cleland},\ and\
  \citenamefont {Martinis}}]{Kelly:2015}%
  \BibitemOpen
  \bibfield  {author} {\bibinfo {author} {\bibfnamefont {J.}~\bibnamefont
  {Kelly}}, \bibinfo {author} {\bibfnamefont {R.}~\bibnamefont {Barends}},
  \bibinfo {author} {\bibfnamefont {A.~G.}\ \bibnamefont {Fowler}}, \bibinfo
  {author} {\bibfnamefont {A.}~\bibnamefont {Megrant}}, \bibinfo {author}
  {\bibfnamefont {E.}~\bibnamefont {Jeffrey}}, \bibinfo {author} {\bibfnamefont
  {T.~C.}\ \bibnamefont {White}}, \bibinfo {author} {\bibfnamefont
  {D.}~\bibnamefont {Sank}}, \bibinfo {author} {\bibfnamefont {J.~Y.}\
  \bibnamefont {Mutus}}, \bibinfo {author} {\bibfnamefont {B.}~\bibnamefont
  {Campbell}}, \bibinfo {author} {\bibfnamefont {Y.}~\bibnamefont {Chen}},
  \bibinfo {author} {\bibfnamefont {Z.}~\bibnamefont {Chen}}, \bibinfo {author}
  {\bibfnamefont {B.}~\bibnamefont {Chiaro}}, \bibinfo {author} {\bibfnamefont
  {A.}~\bibnamefont {Dunsworth}}, \bibinfo {author} {\bibfnamefont {I.-C.}\
  \bibnamefont {Hoi}}, \bibinfo {author} {\bibfnamefont {C.}~\bibnamefont
  {Neill}}, \bibinfo {author} {\bibfnamefont {P.~J.}\ \bibnamefont {O'Malley}},
  \bibinfo {author} {\bibfnamefont {C.}~\bibnamefont {Quintana}}, \bibinfo
  {author} {\bibfnamefont {P.}~\bibnamefont {Roushan}}, \bibinfo {author}
  {\bibfnamefont {A.}~\bibnamefont {Vainsencher}}, \bibinfo {author}
  {\bibfnamefont {J.}~\bibnamefont {Wenner}}, \bibinfo {author} {\bibfnamefont
  {A.~N.}\ \bibnamefont {Cleland}}, \ and\ \bibinfo {author} {\bibfnamefont
  {J.~M.}\ \bibnamefont {Martinis}},\ }\bibfield  {title} {\enquote {\bibinfo
  {title} {State preservation by repetitive error detection in a
  superconducting quantum circuit},}\ }\href {\doibase 10.1038/nature14270}
  {\bibfield  {journal} {\bibinfo  {journal} {Nature}\ }\textbf {\bibinfo
  {volume} {519}},\ \bibinfo {pages} {66--69} (\bibinfo {year}
  {2015})}\BibitemShut {NoStop}%
\bibitem [{\citenamefont {Martinis}(2015)}]{Martinis:2015}%
  \BibitemOpen
  \bibfield  {author} {\bibinfo {author} {\bibfnamefont {J.~M.}\ \bibnamefont
  {Martinis}},\ }\bibfield  {title} {\enquote {\bibinfo {title} {Qubit
  metrology for building a fault-tolerant quantum computer},}\ }\href {\doibase
  10.1038/npjqi.2015.5} {\bibfield  {journal} {\bibinfo  {journal} {npj Quantum
  Information}\ }\textbf {\bibinfo {volume} {1}},\ \bibinfo {pages} {15005}
  (\bibinfo {year} {2015})}\BibitemShut {NoStop}%
\bibitem [{\citenamefont {Jones}\ \emph {et~al.}(2012)\citenamefont {Jones},
  \citenamefont {Meter}, \citenamefont {Fowler}, \citenamefont {McMahon},
  \citenamefont {Kim}, \citenamefont {Ladd},\ and\ \citenamefont
  {Yamamoto}}]{Cody:2012}%
  \BibitemOpen
  \bibfield  {author} {\bibinfo {author} {\bibfnamefont {N.~C.}\ \bibnamefont
  {Jones}}, \bibinfo {author} {\bibfnamefont {R.~V.}\ \bibnamefont {Meter}},
  \bibinfo {author} {\bibfnamefont {A.~G.}\ \bibnamefont {Fowler}}, \bibinfo
  {author} {\bibfnamefont {P.~L.}\ \bibnamefont {McMahon}}, \bibinfo {author}
  {\bibfnamefont {J.}~\bibnamefont {Kim}}, \bibinfo {author} {\bibfnamefont
  {T.~D.}\ \bibnamefont {Ladd}}, \ and\ \bibinfo {author} {\bibfnamefont
  {Y.}~\bibnamefont {Yamamoto}},\ }\bibfield  {title} {\enquote {\bibinfo
  {title} {Layered architecture for quantum computing},}\ }\href {\doibase
  10.1103/PhysRevX.2.031007} {\bibfield  {journal} {\bibinfo  {journal} {Phys.
  Rev. {X}}\ }\textbf {\bibinfo {volume} {2}},\ \bibinfo {pages} {031007}
  (\bibinfo {year} {2012})}\BibitemShut {NoStop}%
\bibitem [{\citenamefont {Monroe}\ and\ \citenamefont
  {Kim}(2013)}]{Monroe:2013}%
  \BibitemOpen
  \bibfield  {author} {\bibinfo {author} {\bibfnamefont {C.}~\bibnamefont
  {Monroe}}\ and\ \bibinfo {author} {\bibfnamefont {J.}~\bibnamefont {Kim}},\
  }\bibfield  {title} {\enquote {\bibinfo {title} {Scaling the ion trap quantum
  processor},}\ }\href {\doibase 10.1126/science.1231298} {\bibfield  {journal}
  {\bibinfo  {journal} {Science}\ }\textbf {\bibinfo {volume} {339}},\ \bibinfo
  {pages} {1164--1169} (\bibinfo {year} {2013})}\BibitemShut {NoStop}%
\bibitem [{\citenamefont {O'Gorman}\ \emph {et~al.}(2016)\citenamefont
  {O'Gorman}, \citenamefont {Nickerson}, \citenamefont {Ross}, \citenamefont
  {Morton},\ and\ \citenamefont {Benjamin}}]{OGorman:2016}%
  \BibitemOpen
  \bibfield  {author} {\bibinfo {author} {\bibfnamefont {J.}~\bibnamefont
  {O'Gorman}}, \bibinfo {author} {\bibfnamefont {N.~H.}\ \bibnamefont
  {Nickerson}}, \bibinfo {author} {\bibfnamefont {P.}~\bibnamefont {Ross}},
  \bibinfo {author} {\bibfnamefont {J.~J.}\ \bibnamefont {Morton}}, \ and\
  \bibinfo {author} {\bibfnamefont {S.~C.}\ \bibnamefont {Benjamin}},\
  }\bibfield  {title} {\enquote {\bibinfo {title} {A silicon-based surface code
  quantum computer},}\ }\href {\doibase 10.1038/npjqi.2015.19} {\bibfield
  {journal} {\bibinfo  {journal} {npj Quantum Information}\ }\textbf {\bibinfo
  {volume} {2}},\ \bibinfo {pages} {15019} (\bibinfo {year}
  {2016})}\BibitemShut {NoStop}%
\bibitem [{\citenamefont {Lekitsch}\ \emph {et~al.}(2017)\citenamefont
  {Lekitsch}, \citenamefont {Weidt}, \citenamefont {Fowler}, \citenamefont
  {M{\o}lmer}, \citenamefont {Devitt}, \citenamefont {Wunderlich},\ and\
  \citenamefont {Hensinger}}]{Lekitsch:2017}%
  \BibitemOpen
  \bibfield  {author} {\bibinfo {author} {\bibfnamefont {B.}~\bibnamefont
  {Lekitsch}}, \bibinfo {author} {\bibfnamefont {S.}~\bibnamefont {Weidt}},
  \bibinfo {author} {\bibfnamefont {A.~G.}\ \bibnamefont {Fowler}}, \bibinfo
  {author} {\bibfnamefont {K.}~\bibnamefont {M{\o}lmer}}, \bibinfo {author}
  {\bibfnamefont {S.~J.}\ \bibnamefont {Devitt}}, \bibinfo {author}
  {\bibfnamefont {C.}~\bibnamefont {Wunderlich}}, \ and\ \bibinfo {author}
  {\bibfnamefont {W.~K.}\ \bibnamefont {Hensinger}},\ }\bibfield  {title}
  {\enquote {\bibinfo {title} {Blueprint for a microwave trapped ion quantum
  computer},}\ }\href {\doibase 10.1126/sciadv.1601540} {\bibfield  {journal}
  {\bibinfo  {journal} {Science}\ }\textbf {\bibinfo {volume} {3}},\ \bibinfo
  {pages} {1--11} (\bibinfo {year} {2017})}\BibitemShut {NoStop}%
\bibitem [{\citenamefont {Clarke}\ and\ \citenamefont
  {Wilhelm}(2008)}]{Clarke:2008}%
  \BibitemOpen
  \bibfield  {author} {\bibinfo {author} {\bibfnamefont {J.}~\bibnamefont
  {Clarke}}\ and\ \bibinfo {author} {\bibfnamefont {F.~K.}\ \bibnamefont
  {Wilhelm}},\ }\bibfield  {title} {\enquote {\bibinfo {title} {Superconducting
  quantum bits},}\ }\href {\doibase 10.1038/nature07128} {\bibfield  {journal}
  {\bibinfo  {journal} {Nature}\ }\textbf {\bibinfo {volume} {453}},\ \bibinfo
  {pages} {1031--1042} (\bibinfo {year} {2008})}\BibitemShut {NoStop}%
\bibitem [{\citenamefont {Boixo}\ \emph {et~al.}(2017)\citenamefont {Boixo},
  \citenamefont {Isakov}, \citenamefont {Smelyanskiy}, \citenamefont {Babbush},
  \citenamefont {Ding}, \citenamefont {Jiang}, \citenamefont {Bremner},
  \citenamefont {Martinis},\ and\ \citenamefont {Neven}}]{Boixo:2017}%
  \BibitemOpen
  \bibfield  {author} {\bibinfo {author} {\bibfnamefont {S.}~\bibnamefont
  {Boixo}}, \bibinfo {author} {\bibfnamefont {S.~V.}\ \bibnamefont {Isakov}},
  \bibinfo {author} {\bibfnamefont {V.~N.}\ \bibnamefont {Smelyanskiy}},
  \bibinfo {author} {\bibfnamefont {R.}~\bibnamefont {Babbush}}, \bibinfo
  {author} {\bibfnamefont {N.}~\bibnamefont {Ding}}, \bibinfo {author}
  {\bibfnamefont {Z.}~\bibnamefont {Jiang}}, \bibinfo {author} {\bibfnamefont
  {M.~J.}\ \bibnamefont {Bremner}}, \bibinfo {author} {\bibfnamefont {J.~M.}\
  \bibnamefont {Martinis}}, \ and\ \bibinfo {author} {\bibfnamefont
  {H.}~\bibnamefont {Neven}},\ }\href {https://arxiv.org/abs/1608.00263}
  {\enquote {\bibinfo {title} {Characterizing quantum supremacy in near-term
  devices},}\ } (\bibinfo {year} {2017}),\ \Eprint
  {http://arxiv.org/abs/arXiv:1608.00263} {arXiv:1608.00263} \BibitemShut
  {NoStop}%
\bibitem [{\citenamefont {Gottesman}(2010)}]{Gottesman:2010}%
  \BibitemOpen
  \bibfield  {author} {\bibinfo {author} {\bibfnamefont {D.}~\bibnamefont
  {Gottesman}},\ }\bibfield  {title} {\enquote {\bibinfo {title} {An
  introduction to quantum error correction and fault-tolerant quantum
  computation},}\ }\href {\doibase 10.1090/psapm/068/2762145} {\bibfield
  {journal} {\bibinfo  {journal} {Quantum Information Science and Its
  Contributions to Mathematics, Proceedings of Symposia in Applied
  Mathematics}\ }\textbf {\bibinfo {volume} {68}},\ \bibinfo {pages} {13--58}
  (\bibinfo {year} {2010})}\BibitemShut {NoStop}%
\bibitem [{\citenamefont {Fowler}\ \emph {et~al.}(2012)\citenamefont {Fowler},
  \citenamefont {Mariantoni}, \citenamefont {Martinis},\ and\ \citenamefont
  {Cleland}}]{Fowler:2012}%
  \BibitemOpen
  \bibfield  {author} {\bibinfo {author} {\bibfnamefont {A.~G.}\ \bibnamefont
  {Fowler}}, \bibinfo {author} {\bibfnamefont {M.}~\bibnamefont {Mariantoni}},
  \bibinfo {author} {\bibfnamefont {J.~M.}\ \bibnamefont {Martinis}}, \ and\
  \bibinfo {author} {\bibfnamefont {A.~N.}\ \bibnamefont {Cleland}},\
  }\bibfield  {title} {\enquote {\bibinfo {title} {Surface codes: Towards
  practical large-scale quantum computation},}\ }\href {\doibase
  10.1103/PhysRevA.86.032324} {\bibfield  {journal} {\bibinfo  {journal} {Phys.
  Rev. A}\ }\textbf {\bibinfo {volume} {86}},\ \bibinfo {pages} {032324}
  (\bibinfo {year} {2012})}\BibitemShut {NoStop}%
\bibitem [{\citenamefont {C{\'{o}}rcoles}\ \emph {et~al.}(2015)\citenamefont
  {C{\'{o}}rcoles}, \citenamefont {Magesan}, \citenamefont {Srinivasan},
  \citenamefont {Cross}, \citenamefont {Steffen}, \citenamefont {Gambetta},\
  and\ \citenamefont {Chow}}]{Corcoles:2015}%
  \BibitemOpen
  \bibfield  {author} {\bibinfo {author} {\bibfnamefont {A.~D.}\ \bibnamefont
  {C{\'{o}}rcoles}}, \bibinfo {author} {\bibfnamefont {E.}~\bibnamefont
  {Magesan}}, \bibinfo {author} {\bibfnamefont {S.~J.}\ \bibnamefont
  {Srinivasan}}, \bibinfo {author} {\bibfnamefont {A.~W.}\ \bibnamefont
  {Cross}}, \bibinfo {author} {\bibfnamefont {M.}~\bibnamefont {Steffen}},
  \bibinfo {author} {\bibfnamefont {J.~M.}\ \bibnamefont {Gambetta}}, \ and\
  \bibinfo {author} {\bibfnamefont {J.~M.}\ \bibnamefont {Chow}},\ }\bibfield
  {title} {\enquote {\bibinfo {title} {Demonstration of a quantum error
  detection code using a square lattice of four superconducting qubits},}\
  }\href {\doibase 10.1038/ncomms7979} {\bibfield  {journal} {\bibinfo
  {journal} {Nature Communications}\ }\textbf {\bibinfo {volume} {6}},\
  \bibinfo {pages} {6979} (\bibinfo {year} {2015})}\BibitemShut {NoStop}%
\bibitem [{\citenamefont {Rist{\`{e}}}\ \emph {et~al.}(2015)\citenamefont
  {Rist{\`{e}}}, \citenamefont {Poletto}, \citenamefont {Huang}, \citenamefont
  {Bruno}, \citenamefont {Vesterinen}, \citenamefont {Saira},\ and\
  \citenamefont {DiCarlo}}]{Riste:2015}%
  \BibitemOpen
  \bibfield  {author} {\bibinfo {author} {\bibfnamefont {D.}~\bibnamefont
  {Rist{\`{e}}}}, \bibinfo {author} {\bibfnamefont {S.}~\bibnamefont
  {Poletto}}, \bibinfo {author} {\bibfnamefont {M.-Z.}\ \bibnamefont {Huang}},
  \bibinfo {author} {\bibfnamefont {A.}~\bibnamefont {Bruno}}, \bibinfo
  {author} {\bibfnamefont {V.}~\bibnamefont {Vesterinen}}, \bibinfo {author}
  {\bibfnamefont {O.-P.}\ \bibnamefont {Saira}}, \ and\ \bibinfo {author}
  {\bibfnamefont {L.}~\bibnamefont {DiCarlo}},\ }\bibfield  {title} {\enquote
  {\bibinfo {title} {Detecting bit-flip errors in a logical qubit using
  stabilizer measurements},}\ }\href {\doibase 10.1038/ncomms7983} {\bibfield
  {journal} {\bibinfo  {journal} {Nature Communications}\ }\textbf {\bibinfo
  {volume} {6}},\ \bibinfo {pages} {6983} (\bibinfo {year} {2015})}\BibitemShut
  {NoStop}%
\bibitem [{\citenamefont {Ofek}\ \emph {et~al.}(2016)\citenamefont {Ofek},
  \citenamefont {Petrenko}, \citenamefont {Heeres}, \citenamefont {Reinhold},
  \citenamefont {Leghtas}, \citenamefont {Vlastakis}, \citenamefont {Liu},
  \citenamefont {Frunzio}, \citenamefont {Girvin}, \citenamefont {Jiang},
  \citenamefont {Mirrahimi}, \citenamefont {Devoret},\ and\ \citenamefont
  {Schoelkopf}}]{Ofek:2016}%
  \BibitemOpen
  \bibfield  {author} {\bibinfo {author} {\bibfnamefont {N.}~\bibnamefont
  {Ofek}}, \bibinfo {author} {\bibfnamefont {A.}~\bibnamefont {Petrenko}},
  \bibinfo {author} {\bibfnamefont {R.}~\bibnamefont {Heeres}}, \bibinfo
  {author} {\bibfnamefont {P.}~\bibnamefont {Reinhold}}, \bibinfo {author}
  {\bibfnamefont {Z.}~\bibnamefont {Leghtas}}, \bibinfo {author} {\bibfnamefont
  {B.}~\bibnamefont {Vlastakis}}, \bibinfo {author} {\bibfnamefont
  {Y.}~\bibnamefont {Liu}}, \bibinfo {author} {\bibfnamefont {L.}~\bibnamefont
  {Frunzio}}, \bibinfo {author} {\bibfnamefont {S.~M.}\ \bibnamefont {Girvin}},
  \bibinfo {author} {\bibfnamefont {L.}~\bibnamefont {Jiang}}, \bibinfo
  {author} {\bibfnamefont {M.}~\bibnamefont {Mirrahimi}}, \bibinfo {author}
  {\bibfnamefont {M.~H.}\ \bibnamefont {Devoret}}, \ and\ \bibinfo {author}
  {\bibfnamefont {R.~J.}\ \bibnamefont {Schoelkopf}},\ }\bibfield  {title}
  {\enquote {\bibinfo {title} {Extending the lifetime of a quantum bit with
  error correction in superconducting circuits},}\ }\href {\doibase
  10.1038/nature18949} {\bibfield  {journal} {\bibinfo  {journal} {Nature}\
  }\textbf {\bibinfo {volume} {536}},\ \bibinfo {pages} {441--445} (\bibinfo
  {year} {2016})}\BibitemShut {NoStop}%
\bibitem [{\citenamefont {Brecht}\ \emph {et~al.}(2016)\citenamefont {Brecht},
  \citenamefont {Pfaff}, \citenamefont {Wang}, \citenamefont {Chu},
  \citenamefont {Frunzio}, \citenamefont {Devoret},\ and\ \citenamefont
  {Schoelkopf}}]{Brecht:2016}%
  \BibitemOpen
  \bibfield  {author} {\bibinfo {author} {\bibfnamefont {T.}~\bibnamefont
  {Brecht}}, \bibinfo {author} {\bibfnamefont {W.}~\bibnamefont {Pfaff}},
  \bibinfo {author} {\bibfnamefont {C.}~\bibnamefont {Wang}}, \bibinfo {author}
  {\bibfnamefont {Y.}~\bibnamefont {Chu}}, \bibinfo {author} {\bibfnamefont
  {L.}~\bibnamefont {Frunzio}}, \bibinfo {author} {\bibfnamefont {M.~H.}\
  \bibnamefont {Devoret}}, \ and\ \bibinfo {author} {\bibfnamefont {R.~J.}\
  \bibnamefont {Schoelkopf}},\ }\bibfield  {title} {\enquote {\bibinfo {title}
  {Multilayer microwave integrated quantum circuits for scalable quantum
  computing},}\ }\href {\doibase 10.1038/npjqi.2016.2} {\bibfield  {journal}
  {\bibinfo  {journal} {npj Quantum Information}\ }\textbf {\bibinfo {volume}
  {2}},\ \bibinfo {pages} {16002} (\bibinfo {year} {2016})}\BibitemShut
  {NoStop}%
\bibitem [{\citenamefont {Brecht}\ \emph {et~al.}(2015)\citenamefont {Brecht},
  \citenamefont {Reagor}, \citenamefont {Chu}, \citenamefont {Pfaff},
  \citenamefont {Wang}, \citenamefont {Frunzio}, \citenamefont {Devoret},\ and\
  \citenamefont {Schoelkopf}}]{Brecht:2015}%
  \BibitemOpen
  \bibfield  {author} {\bibinfo {author} {\bibfnamefont {T.}~\bibnamefont
  {Brecht}}, \bibinfo {author} {\bibfnamefont {M.}~\bibnamefont {Reagor}},
  \bibinfo {author} {\bibfnamefont {Y.}~\bibnamefont {Chu}}, \bibinfo {author}
  {\bibfnamefont {W.}~\bibnamefont {Pfaff}}, \bibinfo {author} {\bibfnamefont
  {C.}~\bibnamefont {Wang}}, \bibinfo {author} {\bibfnamefont {L.}~\bibnamefont
  {Frunzio}}, \bibinfo {author} {\bibfnamefont {M.~H.}\ \bibnamefont
  {Devoret}}, \ and\ \bibinfo {author} {\bibfnamefont {R.~J.}\ \bibnamefont
  {Schoelkopf}},\ }\bibfield  {title} {\enquote {\bibinfo {title}
  {Demonstration of superconducting micromachined cavities},}\ }\href {\doibase
  10.1063/1.4935541} {\bibfield  {journal} {\bibinfo  {journal} {Appl. Phys.
  Lett.}\ }\textbf {\bibinfo {volume} {107}},\ \bibinfo {pages} {192603}
  (\bibinfo {year} {2015})}\BibitemShut {NoStop}%
\bibitem [{\citenamefont {Brecht}\ \emph {et~al.}(2017)\citenamefont {Brecht},
  \citenamefont {Chu}, \citenamefont {Axline}, \citenamefont {Pfaff},
  \citenamefont {Blumoff}, \citenamefont {Chou}, \citenamefont {Krayzman},
  \citenamefont {Frunzio},\ and\ \citenamefont {Schoelkopf}}]{Brecht:2017}%
  \BibitemOpen
  \bibfield  {author} {\bibinfo {author} {\bibfnamefont {T.}~\bibnamefont
  {Brecht}}, \bibinfo {author} {\bibfnamefont {Y.}~\bibnamefont {Chu}},
  \bibinfo {author} {\bibfnamefont {C.}~\bibnamefont {Axline}}, \bibinfo
  {author} {\bibfnamefont {W.}~\bibnamefont {Pfaff}}, \bibinfo {author}
  {\bibfnamefont {J.~Z.}\ \bibnamefont {Blumoff}}, \bibinfo {author}
  {\bibfnamefont {K.}~\bibnamefont {Chou}}, \bibinfo {author} {\bibfnamefont
  {L.}~\bibnamefont {Krayzman}}, \bibinfo {author} {\bibfnamefont
  {L.}~\bibnamefont {Frunzio}}, \ and\ \bibinfo {author} {\bibfnamefont
  {R.~J.}\ \bibnamefont {Schoelkopf}},\ }\href
  {https://arxiv.org/abs/1611.02166} {\enquote {\bibinfo {title} {Micromachined
  integrated quantum circuit containing a superconducting qubit},}\ } (\bibinfo
  {year} {2017}),\ \Eprint {http://arxiv.org/abs/arXiv:1611.02166}
  {arXiv:1611.02166} \BibitemShut {NoStop}%
\bibitem [{\citenamefont {O'Brien}\ \emph {et~al.}(2017)\citenamefont
  {O'Brien}, \citenamefont {Bestwick}, \citenamefont {Vahidpour}, \citenamefont
  {Whyland}, \citenamefont {Angeles}, \citenamefont {Scarabelli}, \citenamefont
  {Villiers}, \citenamefont {Curtis}, \citenamefont {Polloreno}, \citenamefont
  {Selvanayagam}, \citenamefont {Papageorge}, \citenamefont {Rubin},\ and\
  \citenamefont {Rigetti}}]{OBrien:2017}%
  \BibitemOpen
  \bibfield  {author} {\bibinfo {author} {\bibfnamefont {W.}~\bibnamefont
  {O'Brien}}, \bibinfo {author} {\bibfnamefont {A.}~\bibnamefont {Bestwick}},
  \bibinfo {author} {\bibfnamefont {M.}~\bibnamefont {Vahidpour}}, \bibinfo
  {author} {\bibfnamefont {J.~T.}\ \bibnamefont {Whyland}}, \bibinfo {author}
  {\bibfnamefont {J.}~\bibnamefont {Angeles}}, \bibinfo {author} {\bibfnamefont
  {D.}~\bibnamefont {Scarabelli}}, \bibinfo {author} {\bibfnamefont
  {M.}~\bibnamefont {Villiers}}, \bibinfo {author} {\bibfnamefont
  {M.}~\bibnamefont {Curtis}}, \bibinfo {author} {\bibfnamefont
  {A.}~\bibnamefont {Polloreno}}, \bibinfo {author} {\bibfnamefont
  {M.}~\bibnamefont {Selvanayagam}}, \bibinfo {author} {\bibfnamefont
  {A.}~\bibnamefont {Papageorge}}, \bibinfo {author} {\bibfnamefont
  {N.}~\bibnamefont {Rubin}}, \ and\ \bibinfo {author} {\bibfnamefont
  {C.}~\bibnamefont {Rigetti}},\ }\href
  {http://meetings.aps.org/Meeting/MAR17/Session/C46.11} {\enquote {\bibinfo
  {title} {Engineering signal integrity in multi-qubit devices: Part i},}\ }
  (\bibinfo {year} {2017}),\ \Eprint {http://arxiv.org/abs/{C46}.00011}
  {{C46}.00011} \BibitemShut {NoStop}%
\bibitem [{\citenamefont {Rosenberg}\ \emph {et~al.}(2017)\citenamefont
  {Rosenberg}, \citenamefont {Kim}, \citenamefont {Yost}, \citenamefont
  {Mallek}, \citenamefont {Yoder}, \citenamefont {Das}, \citenamefont {Racz},
  \citenamefont {Hover}, \citenamefont {Weber}, \citenamefont {Kerman},\ and\
  \citenamefont {Oliver}}]{Rosenberg:2017}%
  \BibitemOpen
  \bibfield  {author} {\bibinfo {author} {\bibfnamefont {D.}~\bibnamefont
  {Rosenberg}}, \bibinfo {author} {\bibfnamefont {D.}~\bibnamefont {Kim}},
  \bibinfo {author} {\bibfnamefont {D.-R.}\ \bibnamefont {Yost}}, \bibinfo
  {author} {\bibfnamefont {J.}~\bibnamefont {Mallek}}, \bibinfo {author}
  {\bibfnamefont {J.}~\bibnamefont {Yoder}}, \bibinfo {author} {\bibfnamefont
  {R.}~\bibnamefont {Das}}, \bibinfo {author} {\bibfnamefont {L.}~\bibnamefont
  {Racz}}, \bibinfo {author} {\bibfnamefont {D.}~\bibnamefont {Hover}},
  \bibinfo {author} {\bibfnamefont {S.}~\bibnamefont {Weber}}, \bibinfo
  {author} {\bibfnamefont {A.}~\bibnamefont {Kerman}}, \ and\ \bibinfo {author}
  {\bibfnamefont {W.}~\bibnamefont {Oliver}},\ }\href
  {http://meetings.aps.org/Meeting/MAR17/Session/H46.1} {\enquote {\bibinfo
  {title} {{3D} integration for superconducting qubits},}\ } (\bibinfo {year}
  {2017}),\ \Eprint {http://arxiv.org/abs/{H46}.00001} {{H46}.00001}
  \BibitemShut {NoStop}%
\bibitem [{\citenamefont {Kelly}\ \emph {et~al.}(2017)\citenamefont {Kelly},
  \citenamefont {Mutus}, \citenamefont {Lucero}, \citenamefont {Foxen},
  \citenamefont {Graff}, \citenamefont {Klimov}, \citenamefont {Chen},
  \citenamefont {Chiaro}, \citenamefont {Dunsworth}, \citenamefont {Neill},
  \citenamefont {Quintana}, \citenamefont {Wenner},\ and\ \citenamefont
  {Martinis}}]{Kelly:2017}%
  \BibitemOpen
  \bibfield  {author} {\bibinfo {author} {\bibfnamefont {J.}~\bibnamefont
  {Kelly}}, \bibinfo {author} {\bibfnamefont {J.}~\bibnamefont {Mutus}},
  \bibinfo {author} {\bibfnamefont {E.}~\bibnamefont {Lucero}}, \bibinfo
  {author} {\bibfnamefont {B.}~\bibnamefont {Foxen}}, \bibinfo {author}
  {\bibfnamefont {R.}~\bibnamefont {Graff}}, \bibinfo {author} {\bibfnamefont
  {P.}~\bibnamefont {Klimov}}, \bibinfo {author} {\bibfnamefont
  {Z.}~\bibnamefont {Chen}}, \bibinfo {author} {\bibfnamefont {B.}~\bibnamefont
  {Chiaro}}, \bibinfo {author} {\bibfnamefont {A.}~\bibnamefont {Dunsworth}},
  \bibinfo {author} {\bibfnamefont {C.}~\bibnamefont {Neill}}, \bibinfo
  {author} {\bibfnamefont {C.}~\bibnamefont {Quintana}}, \bibinfo {author}
  {\bibfnamefont {J.}~\bibnamefont {Wenner}}, \ and\ \bibinfo {author}
  {\bibfnamefont {J.~M.}\ \bibnamefont {Martinis}},\ }\href
  {http://meetings.aps.org/Meeting/MAR17/Session/H46.7} {\enquote {\bibinfo
  {title} {{3D} integration of superconducting qubits with bump bonds: Part
  2},}\ } (\bibinfo {year} {2017}),\ \Eprint {http://arxiv.org/abs/{H46}.00007}
  {{H46}.00007} \BibitemShut {NoStop}%
\bibitem [{\citenamefont {Chen}\ \emph {et~al.}(2014)\citenamefont {Chen},
  \citenamefont {Megrant}, \citenamefont {Kelly}, \citenamefont {Barends},
  \citenamefont {Bochmann}, \citenamefont {Chen}, \citenamefont {Chiaro},
  \citenamefont {Dunsworth}, \citenamefont {Jeffrey}, \citenamefont {Mutus},
  \citenamefont {O'Malley}, \citenamefont {Neill}, \citenamefont {Roushan},
  \citenamefont {Sank}, \citenamefont {Vainsencher}, \citenamefont {Wenner},
  \citenamefont {White}, \citenamefont {Cleland},\ and\ \citenamefont
  {Martinis}}]{Chen:2014}%
  \BibitemOpen
  \bibfield  {author} {\bibinfo {author} {\bibfnamefont {Z.}~\bibnamefont
  {Chen}}, \bibinfo {author} {\bibfnamefont {A.}~\bibnamefont {Megrant}},
  \bibinfo {author} {\bibfnamefont {J.}~\bibnamefont {Kelly}}, \bibinfo
  {author} {\bibfnamefont {R.}~\bibnamefont {Barends}}, \bibinfo {author}
  {\bibfnamefont {J.}~\bibnamefont {Bochmann}}, \bibinfo {author}
  {\bibfnamefont {Y.}~\bibnamefont {Chen}}, \bibinfo {author} {\bibfnamefont
  {B.}~\bibnamefont {Chiaro}}, \bibinfo {author} {\bibfnamefont
  {A.}~\bibnamefont {Dunsworth}}, \bibinfo {author} {\bibfnamefont
  {E.}~\bibnamefont {Jeffrey}}, \bibinfo {author} {\bibfnamefont {J.~Y.}\
  \bibnamefont {Mutus}}, \bibinfo {author} {\bibfnamefont {P.~J.~J.}\
  \bibnamefont {O'Malley}}, \bibinfo {author} {\bibfnamefont {C.}~\bibnamefont
  {Neill}}, \bibinfo {author} {\bibfnamefont {P.}~\bibnamefont {Roushan}},
  \bibinfo {author} {\bibfnamefont {D.}~\bibnamefont {Sank}}, \bibinfo {author}
  {\bibfnamefont {A.}~\bibnamefont {Vainsencher}}, \bibinfo {author}
  {\bibfnamefont {J.}~\bibnamefont {Wenner}}, \bibinfo {author} {\bibfnamefont
  {T.~C.}\ \bibnamefont {White}}, \bibinfo {author} {\bibfnamefont {A.~N.}\
  \bibnamefont {Cleland}}, \ and\ \bibinfo {author} {\bibfnamefont {J.~M.}\
  \bibnamefont {Martinis}},\ }\bibfield  {title} {\enquote {\bibinfo {title}
  {Fabrication and characterization of aluminum airbridges for superconducting
  microwave circuits},}\ }\href {\doibase 10.1063/1.4863745} {\bibfield
  {journal} {\bibinfo  {journal} {Applied Physics Letters}\ }\textbf {\bibinfo
  {volume} {104}},\ \bibinfo {pages} {052602} (\bibinfo {year}
  {2014})}\BibitemShut {NoStop}%
\bibitem [{\citenamefont {Versluis}\ \emph {et~al.}(2016)\citenamefont
  {Versluis}, \citenamefont {Poletto}, \citenamefont {Khammassi}, \citenamefont
  {Haider}, \citenamefont {Michalak}, \citenamefont {Bruno}, \citenamefont
  {Bertels},\ and\ \citenamefont {DiCarlo}}]{Versluis:2016}%
  \BibitemOpen
  \bibfield  {author} {\bibinfo {author} {\bibfnamefont {R.}~\bibnamefont
  {Versluis}}, \bibinfo {author} {\bibfnamefont {S.}~\bibnamefont {Poletto}},
  \bibinfo {author} {\bibfnamefont {N.}~\bibnamefont {Khammassi}}, \bibinfo
  {author} {\bibfnamefont {N.}~\bibnamefont {Haider}}, \bibinfo {author}
  {\bibfnamefont {D.~J.}\ \bibnamefont {Michalak}}, \bibinfo {author}
  {\bibfnamefont {A.}~\bibnamefont {Bruno}}, \bibinfo {author} {\bibfnamefont
  {K.}~\bibnamefont {Bertels}}, \ and\ \bibinfo {author} {\bibfnamefont
  {L.}~\bibnamefont {DiCarlo}},\ }\href {https://arxiv.org/abs/1612.08208}
  {\enquote {\bibinfo {title} {Scalable quantum circuit and control for a
  superconducting surface code},}\ } (\bibinfo {year} {2016}),\ \Eprint
  {http://arxiv.org/abs/arXiv:1612.08208} {arXiv:1612.08208} \BibitemShut
  {NoStop}%
\bibitem [{\citenamefont {B{\'{e}}janin}\ \emph {et~al.}(2016)\citenamefont
  {B{\'{e}}janin}, \citenamefont {McConkey}, \citenamefont {Rinehart},
  \citenamefont {Earnest}, \citenamefont {McRae}, \citenamefont {Shiri},
  \citenamefont {Bateman}, \citenamefont {Rohanizadegan}, \citenamefont
  {Penava}, \citenamefont {Breul}, \citenamefont {Royak}, \citenamefont
  {Zapatka}, \citenamefont {Fowler},\ and\ \citenamefont
  {Mariantoni}}]{Bejanin:2016}%
  \BibitemOpen
  \bibfield  {author} {\bibinfo {author} {\bibfnamefont {J.}~\bibnamefont
  {B{\'{e}}janin}}, \bibinfo {author} {\bibfnamefont {T.}~\bibnamefont
  {McConkey}}, \bibinfo {author} {\bibfnamefont {J.}~\bibnamefont {Rinehart}},
  \bibinfo {author} {\bibfnamefont {C.}~\bibnamefont {Earnest}}, \bibinfo
  {author} {\bibfnamefont {C.}~\bibnamefont {McRae}}, \bibinfo {author}
  {\bibfnamefont {D.}~\bibnamefont {Shiri}}, \bibinfo {author} {\bibfnamefont
  {J.}~\bibnamefont {Bateman}}, \bibinfo {author} {\bibfnamefont
  {Y.}~\bibnamefont {Rohanizadegan}}, \bibinfo {author} {\bibfnamefont
  {B.}~\bibnamefont {Penava}}, \bibinfo {author} {\bibfnamefont
  {P.}~\bibnamefont {Breul}}, \bibinfo {author} {\bibfnamefont
  {S.}~\bibnamefont {Royak}}, \bibinfo {author} {\bibfnamefont
  {M.}~\bibnamefont {Zapatka}}, \bibinfo {author} {\bibfnamefont
  {A.}~\bibnamefont {Fowler}}, \ and\ \bibinfo {author} {\bibfnamefont
  {M.}~\bibnamefont {Mariantoni}},\ }\bibfield  {title} {\enquote {\bibinfo
  {title} {Three-dimensional wiring for extensible quantum computing: The
  quantum socket},}\ }\href {\doibase 10.1103/PhysRevApplied.6.044010}
  {\bibfield  {journal} {\bibinfo  {journal} {Physical Review Applied}\
  }\textbf {\bibinfo {volume} {6}} (\bibinfo {year} {2016}),\
  10.1103/PhysRevApplied.6.044010}\BibitemShut {NoStop}%
\bibitem [{Note1()}]{Note1}%
  \BibitemOpen
  \bibinfo {note} {We find that Transene~A works well to etch indium thin films
  as well as aluminum.}\BibitemShut {Stop}%
\bibitem [{\citenamefont {Schoeller}\ \emph {et~al.}(2006)\citenamefont
  {Schoeller}, \citenamefont {Kim}, \citenamefont {Park},\ and\ \citenamefont
  {Cho}}]{Schoeller:2006}%
  \BibitemOpen
  \bibfield  {author} {\bibinfo {author} {\bibfnamefont {H.}~\bibnamefont
  {Schoeller}}, \bibinfo {author} {\bibfnamefont {J.}~\bibnamefont {Kim}},
  \bibinfo {author} {\bibfnamefont {S.}~\bibnamefont {Park}}, \ and\ \bibinfo
  {author} {\bibfnamefont {J.}~\bibnamefont {Cho}},\ }\bibfield  {title}
  {\enquote {\bibinfo {title} {Thermodynamics and kinetics of oxidation of pure
  indium solders},}\ }\href {\doibase 10.1557/PROC-0968-V03-07} {\bibfield
  {journal} {\bibinfo  {journal} {{MRS} Proceedings}\ }\textbf {\bibinfo
  {volume} {968}} (\bibinfo {year} {2006}),\
  10.1557/PROC-0968-V03-07}\BibitemShut {NoStop}%
\bibitem [{Note2()}]{Note2}%
  \BibitemOpen
  \bibinfo {note} {This design guarantees that a dc current flows through the
  indium bond and the metallized tunnel when measuring the resistance between
  the two base-chip islands.}\BibitemShut {Stop}%
\bibitem [{Note3()}]{Note3}%
  \BibitemOpen
  \bibinfo {note} {\protect \url
  {http://www.ansys.com/Products/Electronics/ANSYS-Q3D-Extractor}}\BibitemShut
  {NoStop}%
\bibitem [{\citenamefont {Najafi-Yazdi}, \citenamefont {Kelly},\ and\
  \citenamefont {Martinis}(2017)}]{Najafi-Yazdi:2017}%
  \BibitemOpen
  \bibfield  {author} {\bibinfo {author} {\bibfnamefont {A.}~\bibnamefont
  {Najafi-Yazdi}}, \bibinfo {author} {\bibfnamefont {J.}~\bibnamefont {Kelly}},
  \ and\ \bibinfo {author} {\bibfnamefont {J.}~\bibnamefont {Martinis}},\
  }\href {http://meetings.aps.org/Meeting/MAR17/Session/C46.9} {\enquote
  {\bibinfo {title} {High fidelity, numerical investigation of cross talk in a
  multi-qubit xmon processor},}\ } (\bibinfo {year} {2017}),\ \Eprint
  {http://arxiv.org/abs/{C46}.00009} {{C46}.00009} \BibitemShut {NoStop}%
\bibitem [{\citenamefont {Wenner}\ \emph {et~al.}(2011)\citenamefont {Wenner},
  \citenamefont {Neeley}, \citenamefont {Bialczak}, \citenamefont {Lenander},
  \citenamefont {Lucero}, \citenamefont {O'Connell}, \citenamefont {Sank},
  \citenamefont {Wang}, \citenamefont {Weides}, \citenamefont {Cleland},\ and\
  \citenamefont {Martinis}}]{Wenner:2011:a}%
  \BibitemOpen
  \bibfield  {author} {\bibinfo {author} {\bibfnamefont {J.}~\bibnamefont
  {Wenner}}, \bibinfo {author} {\bibfnamefont {M.}~\bibnamefont {Neeley}},
  \bibinfo {author} {\bibfnamefont {R.~C.}\ \bibnamefont {Bialczak}}, \bibinfo
  {author} {\bibfnamefont {M.}~\bibnamefont {Lenander}}, \bibinfo {author}
  {\bibfnamefont {E.}~\bibnamefont {Lucero}}, \bibinfo {author} {\bibfnamefont
  {A.~D.}\ \bibnamefont {O'Connell}}, \bibinfo {author} {\bibfnamefont
  {D.}~\bibnamefont {Sank}}, \bibinfo {author} {\bibfnamefont {H.}~\bibnamefont
  {Wang}}, \bibinfo {author} {\bibfnamefont {M.}~\bibnamefont {Weides}},
  \bibinfo {author} {\bibfnamefont {A.~N.}\ \bibnamefont {Cleland}}, \ and\
  \bibinfo {author} {\bibfnamefont {J.~M.}\ \bibnamefont {Martinis}},\
  }\bibfield  {title} {\enquote {\bibinfo {title} {Wirebond crosstalk and
  cavity modes in large chip mounts for superconducting qubits},}\ }\href
  {\doibase 10.1088/0953-2048/24/6/065001} {\bibfield  {journal} {\bibinfo
  {journal} {Superconductor Science and Technology}\ }\textbf {\bibinfo
  {volume} {24}},\ \bibinfo {pages} {065001} (\bibinfo {year}
  {2011})}\BibitemShut {NoStop}%
\bibitem [{Note4()}]{Note4}%
  \BibitemOpen
  \bibinfo {note} {Five of the nine capped resonators were damaged during
  fabrication and, thus, their parameters are not reported in Table~\ref
  {Table1:McRae:Main}. The fits for the last two capped resonators in the table
  are poor; the corresponding quality factors are not shown.}\BibitemShut
  {Stop}%
\bibitem [{Note5()}]{Note5}%
  \BibitemOpen
  \bibinfo {note} {By measuring a resonator over a time period of ten hours, we
  found a standard deviation as large as half of the mean quality factor.~\cite
  {Neill:2013}}\BibitemShut {NoStop}%
\bibitem [{\citenamefont {Neill}\ \emph {et~al.}(2013)\citenamefont {Neill},
  \citenamefont {Megrant}, \citenamefont {Barends}, \citenamefont {Chen},
  \citenamefont {Chiaro}, \citenamefont {Kelly}, \citenamefont {Mutus},
  \citenamefont {O'Malley}, \citenamefont {Sank}, \citenamefont {Wenner},
  \citenamefont {White}, \citenamefont {Yin}, \citenamefont {Cleland},\ and\
  \citenamefont {Martinis}}]{Neill:2013}%
  \BibitemOpen
  \bibfield  {author} {\bibinfo {author} {\bibfnamefont {C.}~\bibnamefont
  {Neill}}, \bibinfo {author} {\bibfnamefont {A.}~\bibnamefont {Megrant}},
  \bibinfo {author} {\bibfnamefont {R.}~\bibnamefont {Barends}}, \bibinfo
  {author} {\bibfnamefont {Y.}~\bibnamefont {Chen}}, \bibinfo {author}
  {\bibfnamefont {B.}~\bibnamefont {Chiaro}}, \bibinfo {author} {\bibfnamefont
  {J.}~\bibnamefont {Kelly}}, \bibinfo {author} {\bibfnamefont {J.~Y.}\
  \bibnamefont {Mutus}}, \bibinfo {author} {\bibfnamefont {P.~J.~J.}\
  \bibnamefont {O'Malley}}, \bibinfo {author} {\bibfnamefont {D.}~\bibnamefont
  {Sank}}, \bibinfo {author} {\bibfnamefont {J.}~\bibnamefont {Wenner}},
  \bibinfo {author} {\bibfnamefont {T.~C.}\ \bibnamefont {White}}, \bibinfo
  {author} {\bibfnamefont {Y.}~\bibnamefont {Yin}}, \bibinfo {author}
  {\bibfnamefont {A.~N.}\ \bibnamefont {Cleland}}, \ and\ \bibinfo {author}
  {\bibfnamefont {J.~M.}\ \bibnamefont {Martinis}},\ }\bibfield  {title}
  {\enquote {\bibinfo {title} {Fluctuations from edge defects in
  superconducting resonators},}\ }\href {\doibase 10.1063/1.4818710} {\bibfield
   {journal} {\bibinfo  {journal} {Applied Physics Letters}\ }\textbf {\bibinfo
  {volume} {103}},\ \bibinfo {pages} {072601} (\bibinfo {year}
  {2013})}\BibitemShut {NoStop}%
\end{thebibliography}

\begin{thebibliography}{5}%
\makeatletter
\providecommand \@ifxundefined [1]{%
 \@ifx{#1\undefined}
}%
\providecommand \@ifnum [1]{%
 \ifnum #1\expandafter \@firstoftwo
 \else \expandafter \@secondoftwo
 \fi
}%
\providecommand \@ifx [1]{%
 \ifx #1\expandafter \@firstoftwo
 \else \expandafter \@secondoftwo
 \fi
}%
\providecommand \natexlab [1]{#1}%
\providecommand \enquote  [1]{``#1''}%
\providecommand \bibnamefont  [1]{#1}%
\providecommand \bibfnamefont [1]{#1}%
\providecommand \citenamefont [1]{#1}%
\providecommand \href@noop [0]{\@secondoftwo}%
\providecommand \href [0]{\begingroup \@sanitize@url \@href}%
\providecommand \@href[1]{\@@startlink{#1}\@@href}%
\providecommand \@@href[1]{\endgroup#1\@@endlink}%
\providecommand \@sanitize@url [0]{\catcode `\\12\catcode `\$12\catcode
  `\&12\catcode `\#12\catcode `\^12\catcode `\_12\catcode `\%12\relax}%
\providecommand \@@startlink[1]{}%
\providecommand \@@endlink[0]{}%
\providecommand \url  [0]{\begingroup\@sanitize@url \@url }%
\providecommand \@url [1]{\endgroup\@href {#1}{\urlprefix }}%
\providecommand \urlprefix  [0]{URL }%
\providecommand \Eprint [0]{\href }%
\providecommand \doibase [0]{http://dx.doi.org/}%
\providecommand \selectlanguage [0]{\@gobble}%
\providecommand \bibinfo  [0]{\@secondoftwo}%
\providecommand \bibfield  [0]{\@secondoftwo}%
\providecommand \translation [1]{[#1]}%
\providecommand \BibitemOpen [0]{}%
\providecommand \bibitemStop [0]{}%
\providecommand \bibitemNoStop [0]{.\EOS\space}%
\providecommand \EOS [0]{\spacefactor3000\relax}%
\providecommand \BibitemShut  [1]{\csname bibitem#1\endcsname}%
\let\auto@bib@innerbib\@empty
\bibitem [{Note1()}]{Note1}%
  \BibitemOpen
  \bibinfo {note} {A new chamber is being designed made from 316LN stainless
  steel with low magnetic permeability~$\mu \leq 1.005$, reducing magnetic
  contamination of the samples.}\BibitemShut {Stop}%
\bibitem [{\citenamefont {Koppinen}, \citenamefont {V{\"a}ist{\"o}},\ and\
  \citenamefont {Maasilta}(2006)}]{Koppinen:2006}%
  \BibitemOpen
  \bibfield  {author} {\bibinfo {author} {\bibfnamefont {P.~J.}\ \bibnamefont
  {Koppinen}}, \bibinfo {author} {\bibfnamefont {L.~M.}\ \bibnamefont
  {V{\"a}ist{\"o}}}, \ and\ \bibinfo {author} {\bibfnamefont {I.~J.}\
  \bibnamefont {Maasilta}},\ }\bibfield  {title} {\enquote {\bibinfo {title}
  {Effects of annealing to tunnel junction stability},}\ }\href {\doibase
  10.1063/1.2355335} {\bibfield  {journal} {\bibinfo  {journal} {{AIP}
  Conference Proceedings}\ }\textbf {\bibinfo {volume} {850}},\ \bibinfo
  {pages} {1639--1640} (\bibinfo {year} {2006})}\BibitemShut {NoStop}%
\bibitem [{Note2()}]{Note2}%
  \BibitemOpen
  \bibinfo {note} {\protect \url
  {http://www.ansys.com/products/electronics/ansys-hfss}}\BibitemShut {NoStop}%
\bibitem [{\citenamefont {B{\'{e}}janin}\ \emph {et~al.}(2016)\citenamefont
  {B{\'{e}}janin}, \citenamefont {McConkey}, \citenamefont {Rinehart},
  \citenamefont {Earnest}, \citenamefont {McRae}, \citenamefont {Shiri},
  \citenamefont {Bateman}, \citenamefont {Rohanizadegan}, \citenamefont
  {Penava}, \citenamefont {Breul}, \citenamefont {Royak}, \citenamefont
  {Zapatka}, \citenamefont {Fowler},\ and\ \citenamefont
  {Mariantoni}}]{Bejanin:2016}%
  \BibitemOpen
  \bibfield  {author} {\bibinfo {author} {\bibfnamefont {J.}~\bibnamefont
  {B{\'{e}}janin}}, \bibinfo {author} {\bibfnamefont {T.}~\bibnamefont
  {McConkey}}, \bibinfo {author} {\bibfnamefont {J.}~\bibnamefont {Rinehart}},
  \bibinfo {author} {\bibfnamefont {C.}~\bibnamefont {Earnest}}, \bibinfo
  {author} {\bibfnamefont {C.}~\bibnamefont {McRae}}, \bibinfo {author}
  {\bibfnamefont {D.}~\bibnamefont {Shiri}}, \bibinfo {author} {\bibfnamefont
  {J.}~\bibnamefont {Bateman}}, \bibinfo {author} {\bibfnamefont
  {Y.}~\bibnamefont {Rohanizadegan}}, \bibinfo {author} {\bibfnamefont
  {B.}~\bibnamefont {Penava}}, \bibinfo {author} {\bibfnamefont
  {P.}~\bibnamefont {Breul}}, \bibinfo {author} {\bibfnamefont
  {S.}~\bibnamefont {Royak}}, \bibinfo {author} {\bibfnamefont
  {M.}~\bibnamefont {Zapatka}}, \bibinfo {author} {\bibfnamefont
  {A.}~\bibnamefont {Fowler}}, \ and\ \bibinfo {author} {\bibfnamefont
  {M.}~\bibnamefont {Mariantoni}},\ }\bibfield  {title} {\enquote {\bibinfo
  {title} {Three-dimensional wiring for extensible quantum computing: The
  quantum socket},}\ }\href {\doibase 10.1103/PhysRevApplied.6.044010}
  {\bibfield  {journal} {\bibinfo  {journal} {Physical Review Applied}\
  }\textbf {\bibinfo {volume} {6}} (\bibinfo {year} {2016}),\
  10.1103/PhysRevApplied.6.044010}\BibitemShut {NoStop}%
\bibitem [{\citenamefont {Simons}(2001)}]{Simons:2001}%
  \BibitemOpen
  \bibfield  {author} {\bibinfo {author} {\bibfnamefont {R.~N.}\ \bibnamefont
  {Simons}},\ }\href {\doibase 10.1002/0471224758} {\emph {\bibinfo {title}
  {Coplanar Waveguide Circuits, Components, and Systems}}}\ (\bibinfo
  {publisher} {John Wiley {\&} Sons, Inc.},\ \bibinfo {address} {Hoboken, {NJ},
  {USA}},\ \bibinfo {year} {2001})\BibitemShut {NoStop}%
\end{thebibliography}
\end{document}